\documentstyle[epsf,floats,aps,latexsym,twocolumn]{revtex}
\newcommand{\be}{\begin{equation}}
\newcommand{\ee}{\end{equation}}
\newcommand{\bea}{\begin{eqnarray}}  
\newcommand{\eea}{\end{eqnarray}}

\newcommand{\rD}{\mbox{D}}

\newcommand{\p}{\partial}
\newcommand{\s}{\sigma}

\newcommand{\up}{\uparrow}
\newcommand{\down}{\downarrow}
\newcommand{\Up}{\Uparrow}
\newcommand{\Down}{\Downarrow}
\newcommand{\la}{\langle}
\newcommand{\ra}{\rangle}
\newcommand{\rd}{\mbox{d}}
\newcommand{\ri}{\mbox{i}}

\newcommand{\eps}{\epsilon}
\newcommand{\nn}{\nonumber}
\newcommand{\bp}{\bar{\partial}}
\newcommand{\bz}{\bar{z}}
\newcommand{\bw}{\bar{w}}

\newcommand{\vare}{\varepsilon}

\newcommand{\vxi}{\mbox{\boldmath $\xi$}}
\newcommand{\vperp}{\mbox{\boldmath $\perp$}}
\newcommand{\vpi}{\mbox{\boldmath $\pi$}}
\newcommand{\vrho}{\mbox{\boldmath $\rho$}}

\def\tt{{T}}
\def\b{\bar{\beta}}
\def\S{{\cal S}^\alpha}
\def\T{{\cal T}^\alpha}
\begin {document}
\twocolumn[\hsize\textwidth\columnwidth\hsize\csname %
 @twocolumnfalse\endcsname  

\title{\bf Quantum criticalities in a two-leg antiferromagnetic
S=1/2 ladder induced by a staggered magnetic field}
\author{
Y.-J. Wang$~^{a,b}$, F.H.L. Essler$~^c$, M. Fabrizio$~^d$, and
A.A. Nersesyan$~^{a,e}$}
\address{
{$^a$The Abdus Salam International Centre for Theoretical Physics,}
{P.O.Box 586,34100, Trieste, Italy} \\
{$^b$Department of Physics, Nanjing University, 210093, Nanjing, China}\\
{$^c$Department of Physics, University of Warwick, Coventry,}
{CV4 7AL, UK}\\
{$^d$International School for Advanced Studies and INFM, Via Beirut 4,}
{34014, Trieste, Italy}\\
{$^e$The Andronikashvili Insitute of Physics, Tamarashvili 6,}
{390077, Tbilisi, Georgia}
}
\date{\today}
\maketitle

\begin{abstract}
We study a two-leg antiferromagnetic spin-1/2 ladder in the presence of
a staggered magnetic field. We consider two parameter regimes: strong
(weak) coupling along the legs and weak (strong) coupling along the
rungs. In both cases, the staggered field 
drives the Haldane spin-liquid phase of the ladder towards a
Gaussian quantum criticality. In a generalized spin ladder with a non-Haldane,
spontaneously dimerized phase, the staggered magnetic field 
induces an Ising quantum critical
regime. In the vicinity of the critical lines, we derive low-energy
effective field theories and use these descriptions to determine the
dynamical response functions, the staggered spin susceptibility and the
string order parameter. 
\end{abstract}
\medskip
]
\narrowtext

\section{Introduction}

The problem of quantum critical points (QCP) is one of the most important issues in the
physics of strongly correlated electron systems, in particular in the context of
high-$T_c$ superconductivity\cite{sachdev}
and heavy-fermion compounds \cite{continentino}.
Recently this problem attracted much interest also in the context 
of one-dimensional (1D) quantum systems, such as 1D interacting electrons,
antiferromagnetic spin chains and ladders, 
where a detailed description of QCPs
is available due to the powerful nonperturbative techniques based on 
conformal field theory, bosonization and integrability.
It is well known that 
universal properties of 1D quantum systems 
can be described on the basis of a properly chosen conformally invariant theory
deformed by a number of perturbations consistent with the
structure and symmetry of the underlying miscroscopic model. Quantum criticalities
can then emerge due to the competition between two (or more) relevant perturbations
which, when acting alone, would drive the system to
qualitatively different strong-coupling massive phases
that cannot be smoothly connected by a continuous path in the parameter
space of the model. 
An example of a theory of this kind,
displaying an Ising quantum criticality,
is the so-called double-frequency sine-Gordon model \cite{dm,fgn,fgn2}.

In connection with 1D quantum antiferromagnets,
a plausible QCP scenario has been anticipated some time ago by Affleck and Haldane
\cite{ah}. They argued that a massive phase of a translationally invariant spin-chain
Hamiltonian can be pushed towards quantum criticality by an external, parity-breaking
perturbation.
Typical examples of such perturbations are an explicit dimerization,
whose role in the formation of QCPs 
has already been analyzed in several spin-chain and spin-ladder 
models \cite{snake,cg,ts,wn}, and a staggered magnetic
field. While dimerization of quantum spin chains and ladders
is quite realistic  
because it can originate, for instance, from the spin-phonon coupling,
the case of a static magnetic field whose sign alternates on a microscopic scale 
used to be regarded as not achievable in experimental conditions.

However, recently two beautiful experimental realizations of staggered
magnetic fields in quasi-1D magnetic insulators have been discovered. The
first concerns 
the spin-1/2 antiferromagnetic chain compound Copper Benzoate
\cite{dender}. Due to the low crystalline symmetry, the magnetic field $H$
couples to the effective spins-1/2 through a gyromagnetic {\sl tensor}
\cite{gtensor}
\begin{equation}
H_{\rm magn}= \sum_n\sum_{a,b} \left[g_{\alpha \beta}^{\rm u}+(-1)^n\ 
g_{\alpha \beta}^{\rm st}\right]  H_{\alpha}\ S^{\beta}_{n}\ .
\end{equation}
Application of a uniform magnetic field ${\bf H}$ thus induces a
staggered field ${\bf h}$ in a direction perpendicular to ${\bf H}$.
In Copper Benzoate there is a second mechanism that gives rise to
a staggered internal magnetic field.
It derives from the staggered
Dzyaloshinskii-Moriya (DM) interaction along the chain
direction 
\be
{\cal H}_{\rm DM} = \sum_j (-1)^j\
{\bf D}\cdot({\bf S}_j\times{\bf S}_{j+1}), 
\label{hdm}
\ee
which, when a uniform field ${\bf H}$ is applied, induces  
a staggered component proportional to ${\bf H}\times {\bf D}$
\cite{oshikawa1}. The presence of a staggered field has been shown to
lead to a variety of very interesting consequences
\cite{oshikawa1,essler1,essler2,oshikawa2}. 
The staggered field scenario described above is by
no means specific to Copper Benzoate. There are at least two other
materials, ${\rm Yb_4As_3}$ \cite{YAs}
and ${\rm [PM\cdot Cu(NO_3)_2\cdot(H_2O)_2]_n}$ \cite{feyer},
whose properties in a magnetic field are controlled
by the same mechanism.

A completely different mechanism that leads to the generation of a
staggered field  has recently been discovered for spin-1 Haldane gap
compounds of the type ${\rm R_2BaNiO_3}$ \cite{zheludev,maslov}, where
$R$ is a magnetic rare earth. In these materials, the rare-earth ions
are only weakly coupled to the Ni chains, but interact strongly with
one another. They may be considered to live on a separate sublattice
that undergoes a N\'eel transition at a rather high temperature $T_N$.
The effect of the resulting antiferromagnetic order is to induce an
effective staggered magnetic field along the Ni (spin-1 Heisenberg)
chains below $T_N$.

In this paper, we study the effect of an external staggered magnetic field
on the low-energy properties of the spin-1/2 antiferromagnetic two-chain 
Heisenberg ladder. Although at present no analogous mechanism for 
the generation of a staggered
field has been found for Heisenberg ladders compounds, in
principle either of the two scenarios mentioned above are
possible. One may well expect that a staggered field will be realized
in a ladder compound before long, so that addressing this problem is not
only of academic interest.

The Hamiltonian of the ``standard'' ladder is
\bea
H_{\rm stand} &=& J \sum_{j=1,2} \sum_n {\bf S}_{j,n} \cdot {\bf S}_{j,n+1}\nn\\
&+& J_{\perp} \sum_n {\bf S}_{1,n} \cdot {\bf S}_{2,n}
- h \sum_{a=1,2} \sum_n (-1)^n S^z _{a,n},
\label{model}
\eea
where $J$ and $J_{\perp}$ are antiferromagnetic exchange coupling
constants in the ``leg'' and ``rung'' directions, respectively.
We employ weak-coupling ($J_{\perp} \ll J$) and strong-coupling
($J_{\perp} \gg J$) approaches to show that there exists a
critical value of the staggered magnetic field, $h=h_c (J_{\perp},J)$, 
for which the system displays a Gaussian U(1) criticality with central
charge $C=1$, characterized by nonuniversal critical exponents.
Both at $h<h_c$ and $h>h_c$ the spectrum is gapped, and the spin correlations
are commensurate with the underlying lattice. This is diffferent from
the spin ladder in a uniform magnetic field \cite{chitra} which induces a
transition from the gapped commensurate phase ($h<h_c$) to a gapless
incommensurate phase ($h>h_c$).
Comparing the results of the  
weak-coupling and strong-coupling approaches, 
we find that near the critical point the low-energy
properties of the spin ladder are adequately described in
terms of a XXZ spin-1/2 chain with a $J$ and $J_{\perp}$-dependent
exchange anisotropy and an effective staggered magnetic field
proportional to $h - h_c$. Hence we expect that the existence of
the U(1) QCP is a universal property of the standard Heisenberg
spin-1/2 ladder in a staggered field.

The critical surface $h_c (J_{\perp},J)$
separates two massive phases: an anisotropic Haldane spin-liquid
phase at $h < h_c$ with coherent $S^z = \pm 1$ and $S^z = 0$ magnon
excitations having different, field dependent, mass gaps, 
and another massive phase at $h > h_c$
in which the spin excitation spectrum at $q\sim \pi$ still includes
coherent transverse ($S^z = \pm 1$) magnons, whereas the 
$S^z = 0$ modes transform to an incoherent background.
The transition is associated with softening of the $S^z = \pm 1$ spin-doublet modes
and is characterized by a divergent staggered magnetic susceptibility.

We also discuss the properties of a generalized ladder,
\bea
H_{\rm gen} &=& H_{\rm stand} + V \sum_n ({\bf S}_{1,n} \cdot {\bf S}_{1,n+1})
({\bf S}_{2,n} \cdot {\bf S}_{2,n+1})\ ,
\label{gen.model}
\eea
which, apart from the on-rung interchain exchange $J_{\perp}$, also includes
a four-spin interaction $V$. This model is interesting because it can display
non-Haldane, spontaneously dimerized massive phases if
the biquadratic interaction $V$ is sufficiently large \cite{nt}.
The existence of such interactions in ladder compounds is supported by
recent neutron scattering experiments \cite{matsuda}.
We show that, in the weak coupling regime, the staggered magnetic field
can drive the non-Haldane phase to an Ising quantum criticality (with
central charge $C=1/2$), where the spontaneous dimerization vanishes
and the staggered magnetic susceptibility is logarithmically
divergent. In the dimerized phase ($h<h_c$) the spin excitation
spectrum is entirely incoherent, whereas at $h>h_c$, as in the
large-$h$ phase of the standard ladder, the coherent $S^z = \pm 1$
magnons are recovered.

\begin{figure}[ht]
\begin{center}
\epsfxsize=0.45\textwidth
\epsfbox{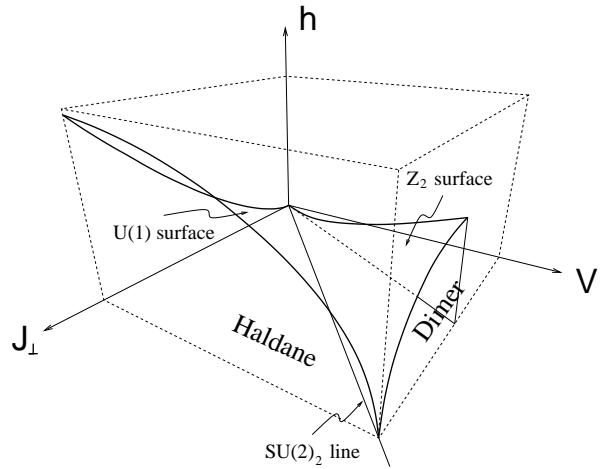}
\end{center}
\caption{
Phase diagram of the generalized spin ladder in
a staggered magnetic field in the limit $J_\perp, V \ll J$. The SU(2)$_2$ critical
line, which at $h=0$ separates the Haldane spin-liquid phase
from the spontaneously dimerized phase, is split by the field
into U(1) and Z$_2$ critical surfaces.}
\label{fig:fig1}
\end{figure}

The qualitative phase diagram in the $(J_\perp,h,V)$ space is 
shown in Fig.\ref{fig:fig1}. We note that the U(1) (Gaussian, $C=1$) and 
Z$_2$ (Ising, $C=1/2$) critical surfaces merge at $h=0$ into a critical 
line,   
which has been shown earlier \cite{nt} 
to belong to the universality
class of the SU(2)$_2$ Wess-Zumino-Novikov-Witten model with central
charge $C=3/2$.

\medskip

The paper is organized as follows. In section 2, we study the phase diagram of
the standard spin-ladder model (\ref{model}) in the weak-coupling
limit ($J_{\perp}, h \ll J$). Here we employ a field-theoretical
approach which represents the low-energy sector of the spin ladder as
an SO(3)$\times$Z$_2$-symmetric model of four noncritical 2D Ising systems\cite{snt,nt}.
Integrating out the fast degrees of freedom associated with collective
singlet excitations, we derive an effective action 
which describes the triplet sector with anisotropy induced by the staggered field.
We demonstrate the existence of a Gaussian criticality and show that, close to the
critical point, the model  
is described in terms of a spin-1/2 XXZ chain
with a small anisotropy parameter and an effective staggered magnetic
field $h^* \propto h-h_c$. In this section we also derive projections of all
physical fields onto the low-energy triplet sector.

In section 3 we discuss general properties of the dynamical structure factor,
measured in neutron scattering experiments, 
for quantum spin chains and ladders in a staggered magnetic field.
In section 4 we determine the dynamical structure factor 
of a weakly coupled spin ladder for momentum
transfers along the ``leg'' direction close to $\pi$ and $0$ in both massive
phases and at criticality
In section 5 we study the induced staggered magnetization
and show that at the U(1) transition the staggered
susceptibility is divergent with a nonuniversal critical exponent. 

In section 6 we consider the model (\ref{model}) in the
strong-coupling limit ($J_{\perp}, h \gg J$). 
Treating the exchange $J$ as a perturbation and projecting
the Hamiltonian onto the subspace of the low energy states, we arrive at an
effective strongly anisoptropic spin-1/2 Heisenberg chain model in a staggered
magnetic field $h^*  = h - J_{\perp} + J/2 + \cdots$. 
As opposed to the weak-coupling case, in the strong-coupling limit
the relationship between the
parameters of the original and the effective low-energy model is known 
with any desired degree of accuracy. 
In section 7 we 
exploit the exact integrability of the
sine-Gordon model and apply the formfactor method
to achieve a quantitative
description of the dynamical properties of the model in the strong coupling 
limit. This section also contains a brief overview of the formfactor appoach.

In section 8, we turn to the generalized ladder (\ref{gen.model}) and
describe the Ising transition in the non-Haldane phase, induced by 
the staggered magnetic field. 
In section 9 we address the topological order of the generalized ladder model
in the staggered field and analyze the field dependence 
of the longitudinal and transverse components of the nonlocal string order 
parameter in various parts of the phase diagram shown in Fig.\ref{fig:fig1}. 

A comparison of 
the results obtained for in the weak-coupling and strong-coupling regimes
and our final conclusions are given in section 10. 
The paper is supplemented with four Appendices which contain some technical 
details used in the main text.

\section{Weak-coupling limit}
\subsection{Ising-model description of the Heisenberg ladder}

In this and next sections we will be concerned with a weakly coupled
spin ladder in a small staggered magnetic field:
$
J_{\perp}, h \ll J.
$
We start our discussion with a brief overview of the effective 
field-theoretical model
describing universal properties of the spin-liquid state in the 
antiferromagnetic two-leg Heisenberg ladder \cite{snt,nt}.

It is well known that a single S=1/2 Heisenberg chain
is critical and has massless S=1/2 spinons as elementary 
excitations \cite{ft}. 
When a small $J_{\perp}$ is switched on, the spinons
of the originally decoupled chains get confined to form gapped triplet
and singlet excitations (for a review see Ref.\cite{dr}). 
The field theory that accounts for the spinon-magnon transmutation in
the two-leg spin-1/2 ladder represents 
an O(3)$\times$Z$_2$-symmetric model
of four massive real (Majorana) fermions, or equivalently,
four noncritical 2D Ising models\cite{snt,nt,gnt}. In the continuum limit, 
the Hamiltonian density
\bea
&&{\cal H}_M =  \sum_{a=1,2,3} \left[ - \frac{\ri v_t}{2}
\left( \xi^a _R \p_x \xi^a _R - \xi^a _L \p_x \xi^a _L \right)
- \ri m_t \xi^a _R \xi^a _L \right] \nn\\
&&-  \frac{\ri v_s}{2}
\left( \xi^4 _R \p_x \xi^4 _R - \xi^4 _L \p_x \xi^4 _L \right)
- \ri m_s \xi^4 _R \xi^4 _L 
+ {\cal H}_{\rm marg}
\label{maj-Ham}
\eea
describes a degenerate triplet of Majorana fields, 
$\vxi_{\nu} = (\xi^1 _{\nu}, \xi^2 _{\nu},\xi^1 _{\nu})$,
$(\nu=R,L)$, 
and a singlet Majorana field $\xi^4 _{R,L}$. 
The velocities $v_{t,s}$ are proportional to $J \alpha$, where
$\alpha$ is a short-distance cutoff of the theory. 
The triplet and singlet masses
are given by 
\be
m_t = J_{\perp} \lambda, ~~~m_s = - 3 J_{\perp} \lambda,
\label{bare-masses} 
\ee
where $\lambda$ is a nonuniversal constant. The corresponding
correlation lengths in the triplet and singlet sectors of the model
are 
\be
l_{t,s} \sim v_{t,s}/|m_{t,s}|, \label{corr.lengths}
\ee
which for a weakly coupled ladder are macroscopically
large ($l_{t,s} \gg \alpha$). The last term in (\ref{maj-Ham})
describes a weak interaction between the Majorana fermions, 
\be
{\cal H}_{\rm marg} =  \frac{1}{2} g_1 \left( \vxi_R \cdot \vxi_L \right)^2
+ g_2 \left( \vxi_R \cdot \vxi_L \right)\left( \xi^4 _R \xi^4 _L \right),
\label{marg.int}
\ee
where $g_1 = - g_2 = \frac{1}{2}\pi \alpha J_{\perp}$.

For the generalized ladder (\ref{gen.model}),
the low-energy effective model is still of the form (\ref{maj-Ham}), (\ref{marg.int}),
with the only difference that the triplet and singlet masses 
\be
m_t = J_{\perp} \lambda  - V \lambda' , 
~~~m_s = - 3 J_{\perp} \lambda - V \lambda'
\label{bare-masses-gen}
\ee
($\lambda'$ is a nonuniversal constant) can be varied independently.

In the continuum description, the local spin densities of the two 
Heisenberg chains, ${\bf S}_{j} (x)~(j=1,2)$, are contributed by low-energy
spin-fluctuation modes centered in the momentum space at $q=0$ and $q=\pi$.
Accordingly
\be
{\bf S}_j (x) = {\bf J}_{jR} (x)  + {\bf J}_{jL} (x)
+ (-1)^{x/a_0} {\bf n}_j (x).
\label{spin-decom}
\ee
The chiral components of the vector currents ${\bf J}_{j;R,L}$ 
(i.e. the smooth parts of the spin densities)
can be expressed locally in terms of the Majorana bilinears
\bea
&&{\bf I}_{\nu} = {\bf J}_{1\nu} + {\bf J}_{2\nu} = - \frac{\ri}{2}
\left( \vxi_{\nu} \wedge \vxi_{\nu} \right), \nn\\
&&{\bf K}_{\nu} = {\bf J}_{1\nu} - {\bf J}_{2\nu} 
= \ri \vxi_{\nu} \xi^4 _{\nu}~~~~~~(\nu = R,L).
\label{vec-currents-maj}
\eea
However, the most strongly fluctuating fields of the spin ladder,
the staggered magnetizations ${\bf n}_j(x)$ and dimerization fields
$\eps_j (x)\longrightarrow (-1)^n {\bf S}_{j,n} \cdot {\bf S}_{j,n+1}$, all
with scaling dimension 1/2, are nonlocal with respect to  $\vxi,
\xi^4$. These fields, however, admit a representation in terms
of the order ($\s$) and disorder ($\mu$) operators of the related
noncritical Ising models \cite{snt,nt,gnt}:
\bea
&&{\bf n}^{+} \sim (1 /\alpha) \left(
\mu_1 \s_2 \s_3 \mu_4, ~\s_1 \mu_2 \s_3 \mu_4, ~ \s_1 \s_2 \mu_3 \mu_4 \right)
\label{n+}\\
&&{\bf n}^{-} \sim  (1 /\alpha) \left(
\s_1 \mu_2 \mu_3 \s_4, ~\mu_1 \s_2 \mu_3 \s_4, ~ \mu_1 \mu_2 \s_3 \s_4 \right)
\label{n-}\\
&&\eps^{+} \sim (1/\alpha)\mu_1 \mu_2 \mu_3 \mu_4, ~~~~
\eps^{-} \sim (1 /\alpha)\s_1 \s_2 \s_3 \s_4,
\label{eps-pm}
\eea
where ${\bf n}^{\pm} = {\bf n}_1 \pm {\bf n}_2$ and
$\eps^{\pm} = \eps_1 \pm \eps_2$.

Since the the correlation lengths $l_{t,s}$ are large, 
all Ising systems are slightly noncritical. Whether they occur in the
ordered or disordered phase depends on the sign of the corresponding mass
$m \propto (T-T_c)/T_c$. 
The crucial property of the standard ladder is that
the signs of the triplet and singlet Majorana masses are always {\sl
opposite}. 
This fact, together with the known asymptotics of
the two-point correlation functions of a noncritical Ising model \cite{wu}, 
leads to the observation that
the dynamical spin susceptibility of the antiferromagnetic Heisenberg ladder,
$\chi'' (q,\omega)$, obtained by Fourier transforming 
the correlation function $\la {\bf n}^- (x,\tau) \cdot {\bf n}^- (0,0) \ra$,
exhibits a coherent S=1 single-magnon peak at $\omega^2 = (\pi - q)^2
v_t ^2 + m^2 _t$ (with $q$ close to $\pi$).
Due to multiparticle processes, the dynamical spin susceptibility
$\chi''(q\approx \pi, \omega)$ also displays an incoherent tail with a 
threshold at $\omega = 3m_t$. The singlet mode shows up only at higher
energies, $\omega > 2m_t + |m_s| \sim 5 m_t$.
Thus, at low energies, the standard two-chain ladder
represents a disordered spin liquid, 
similar to the Haldane phase of the spin-1 Heisenberg chain with a small
triplet gap.

Let us now switch on a small staggered magnetic field 
${\bf h} = h \hat{z}$ which is assumed to be the same for the both chains of the ladder.
The field couples to the total staggered magnetization $n^{+}_z$, and
the Hamiltonian density becomes
\be
{\cal H} = {\cal H}_{M} - (\bar{h}/\alpha) \s_1 \s_2 \mu_3 \mu_4 ,
\label{total-ham-dens}
\ee
where $\bar{h} \sim h$. 
Here a comment is in order.
In spite of the already mentioned similarity between the antiferromagnetic
S=1/2 ladder and the spin-1 chain, 
it would be misleading to think that
the role of the staggered field in these two
cases will be similar. 
A weakly coupled two-leg S=1/2 ladder can be mapped onto a spin-1 chain
by formally shifting the singlet excitation band to infinity. This implies
the substitutions $\mu_4 \to \la \mu_4 \ra = 0$, $\s_4 \to \la \s_4 \ra \neq 0$,
in which case the $q=\pi$ component of the S=1 spin density is determined by
the {\sl relative} staggered magnetization of the spin ladder, 
${\bf n}^-$ [see Eq.(\ref{n-:proj})].
So, for the S=1 chain the magnetic term has a structure different from (\ref{total-ham-dens}):
\be
{\cal H}^{S=1}_{\rm mag} = h \mu_1\mu_2\s_3. \label{mag-term:S=1}
\ee
Since $m_t > 0$,
in the leading order the interaction (\ref{mag-term:S=1}) gives rise to an effective
magnetic field $\tilde{h} \sim h \la\mu_1 \mu_2\ra$ applied to the third Ising system
($\tilde{h} \s_3$). The spectrum of the S=1 chain in a staggered
field is therefore always massive (see Refs.\cite{S1stag}).

The existence of a U(1) transition in the model (\ref{total-ham-dens}) can be foreseen
as follows.
Since the triplet of Ising copies is disordered, 
the magnetic term in (\ref{total-ham-dens})
can be approximately replaced by
$\tilde{h} \s_1 \s_2 \mu_4$, where $\tilde{h} \sim \bar{h} \la \mu_3 \ra$. 
Making a duality transformation in the fourth (singlet)
Ising copy ($m_s \to - m_s$, $\mu_4 \leftrightarrow \s_4$), one arrives at a system of three
disordered Ising models with the underlying U(1)$\times$Z$_2$ symmetry, coupled
by the interaction $\tilde{h} \s_1 \s_2 \s_4$. This is
an Ising-model representation of an anisotropic
spin-1 chain close to the integrable, multicritical point\cite{tb},
with a perturbation representing a parity-breaking, dimerization
field. In the isotropic, SU(2)-symmetric case, this model is
known\cite{snake,cg,ts,wn,s=1.dim} to exhibit a QCP where it becomes
equivalent to the S=1/2 Heisenberg chain (SU(2)$_1$ WZNW universality
class). A finite easy-plane anisotropy transforms this criticality to a Gaussian one.

\subsection{Effective action in the triplet sector}

Turning back to Eq.(\ref{total-ham-dens}), we notice that the magnetic term
contains the disorder operator $\mu_4$ of the singlet Ising system which 
has zero expectation value and represents
``fast'' degrees of freedom of the system
(although the ratio $|m_s|/m_t \simeq 3$ may not seem large enough, it can
be significantly increased in the generalized model (\ref{gen.model})
with the ``spin-liquid'' 
condition $m_s m_t < 0$ still preserved; see Eqs. (\ref{bare-masses-gen})).
We will therefore integrate the singlet mode out to
obtain an effective action in the triplet sector. The existence of a Gaussian
criticality will then immediately follow from the structure of this action.

We write the total Euclidean action of the model as $S = S_t + S_s + S_{st}$, 
where $S_t [\vxi]$ and $S_s [\xi^4]$ are the contributions of triplet and
singlet sectors, respectively, and
\bea
&&S_{st} [\vxi, \xi^4]
= \frac{1}{v} \int \rd^2 {\bf r} \left[ {\cal O}_h ({\bf r}) + 
 {\cal O}_g ({\bf r})\right], \nn\\
&&{\cal O}_h  =  
- (\bar{h}/\alpha) \s_1  \s_2  \mu_3  \mu_4 ,\
{\cal O}_g =  g_2 \left( \vxi_R  \cdot \vxi_L \right)
\left( \xi^4 _R \xi^4 _L  \right)
\label{pertur-action}
\eea  
is treated as a perturbation.
Here ${\bf r} = (x, v\tau)$, and for simplicity we ignore the
difference between the the triplet and singlet velocities. 
Integrating over $\xi^4$ in the
partition function
$$
Z = \int \rD [\vxi] \rD [\xi^4]~ e^{- S[\vxi, \xi^4]}
= {\rm const}~\int \rD [\vec{\xi}]~ e^{- S_{\rm eff} [\vxi]}
$$
yields the effective action in the triplet sector in the form of 
a cumulant expansion:
\be
S_{\rm eff} [\vxi] = S_t [\vxi] + \la S_{st} \ra_s
- \frac{1}{2} \left[ \la S^2 _{st} \ra_s - \la S_{st} \ra^2 _s  \right]
+ \cdots ,
\label{eff-action}
\ee
where $\la \cdots \ra_s$ means averaging over the free massive fermions $\xi^4$.

The first-order correction in the expansion (\ref{eff-action}) gives rise
to a small renormalization of the triplet mass
$$
m_t \to m_t + m_s \frac{g_2}{2\pi v} \ln \left(\frac{v}{|m_s| \alpha}\right).
$$
The cross term proportional to $g_2 h$ involves the correlator 
$
\la \mu_4 ({\bf r}_1) \xi^4 _R ({\bf r}_2)\xi^4 _L ({\bf r}_2)
\ra_s
$
which vanishes due to the unbroken ($\mu \to - \mu$, $\xi_{R,L} \to - \xi_{R,L}$) 
symmetry of an ordered Ising model (see Appendix A). 
In the second order in $g_2$, one obtains terms leading to
renormalization of 
the 
velocity $v_t$ and coupling constant $g_1$.
Assuming that 
all these renormalizations are already taken into account, we are left with the following
expression for the effective action:
\bea
&&S_{\rm eff} [\vxi] = S_t [\vxi]\nn\\
&&- \frac{{\bar h}^2}{2v^2 \alpha^2} \int \rd^2 {\bf r}_1\rd^2 {\bf r}_2
~{\cal O}_t ({\bf r}_1){\cal O}_t ({\bf r}_2)~
\la \mu_4 ({\bf r}_1)\mu_4 ({\bf r}_2) \ra_s,
\label{eff.action2}
\eea
where ${\cal O}_t = \s_1 \s_2 \mu_3$.

The fourth Ising copy is ordered ($m_s < 0$); so the correlation
function in (\ref{eff.action2}) 
decays exponentionally at distances
$r\sim l_s$ \cite{wu}
\be
\la \mu_4 ({\bf r})\mu_4 ({\bf 0}) \ra_s 
\simeq \frac{A_1 (\alpha/ l_s)^{1/4}}{\sqrt{2\pi r/l_s}} e^{- r/l_s} + 
O\left(e^{-3r/l_s}  \right),
\label{mu4-mu4-corr}
\ee
where $A_1$ is a nonuniversal parameter.
Under the assumption that $|m_s|/m_t \gg 1$,
the correlation length in the triplet sector, $l_t$, is much
larger than that in the singlet sector, $l_s$. Therefore the product
${\cal O}_t ({\bf r}_1){\cal O}_t ({\bf r}_2)$ in Eq.(\ref{eff.action2})
can be treated by means of the short-distance operator product expansion (OPE).
Using the fusion rules for two Ising order and disorder operators
(see Appendix A, Eqs. (\ref{ope11}),(\ref{ope12})), we find that
\bea
&&S_{\rm eff} [\vxi] = S_t [\vxi] \nn\\
&&
+~ \frac{\ri C_1}{\alpha} \left(\frac{{\bar h}}{m_s} \right)^2
\int \rd^2 {\bf r} \left( \xi^1 _R \xi^1 _L + \xi^2 _R \xi^2 _L -
\xi^3 _R \xi^3 _L \right)\nn\\
&&+~ C_2 \left(  \frac{l_s}{\alpha}\right)\left(\frac{{\bar h}}{m_s} \right)^2\nn\\
&& \times
\int \rd^2 {\bf r} \left[\xi^1 _R \xi^1 _L \xi^2 _R \xi^2 _L
- \left( \xi^1 _R \xi^1 _L + \xi^2 _R \xi^2 _L \right) \xi^3 _R \xi^3 _L \right],
\label{eff.trip.action}
\eea
where 
$C_1$ and $C_2$ are positive numerical constants.

Thus, we arrive at the following effective Hamiltonian for the triplet 
degrees of freedom
\bea
{\cal H}_{\rm t;eff} &=& 
 - \frac{\ri v}{2} \left( \vxi_R \cdot \p_x \vxi_R
- \vxi_L \cdot \p_x \vxi_L \right) - \ri m_a \xi^a _R \cdot \xi^a _L \nn\\
&+& g_{\parallel} I^3 _R I^3 _L +
g_{\perp} \left(I^1 _R I^1 _L + I^2 _R I^2 _L \right),
\label{eff.trip.ham}
\eea
which has the same structure as the field-theoretical model suggested
by Tsvelik \cite{tsv} to describe the Heisenberg spin--1 chain with a
biquadratic term and a single-ion anisotropy. 
The staggered magnetic field introduces anisotropy in the spin ladder and
effectively lowers
the SO(3) symmetry of the Majorana triplet down to SO(2) $\times$
Z$_2$ by splitting the masses,
\bea
m_d \equiv m_1 &=& m_2 = m_t - C_1 \left( \frac{v}{\alpha} \right) 
\left(\frac{\bar h}{m_s}  \right)^2,\nn\\
m_3 &=& m_t + C_1 \left( \frac{v}{\alpha} \right) 
\left(\frac{\bar h}{m_s}  \right)^2,
\label{renorm.masses}
\eea
and by renormalising the coupling constants of the marginal
current-current interaction
\bea
g_{\parallel} &=& g_1 + C_2 v \left( \frac{l_s}{\alpha} \right)
\left(\frac{\bar h}{|m_s|}  \right)^2,\nn\\
g_{\perp} &=& g_1 - C_2 v \left( \frac{l_s}{\alpha} \right)
\left(\frac{\bar h}{|m_s|}  \right)^2 .
\label{renorm.couplings}
\eea

From (\ref{renorm.masses}) it follows that increasing the staggered
magnetic field increases the mass $m_3$ 
whereas the mass of the Majorana doublet, $m_d$,
 decreases
and vanishes at a critical value of the field
\be
h_c \propto |m_s| \left( \frac{\alpha}{l_t}\right)^{1/2} \propto
J^{3/2}_{\perp}/ J^{1/2}.
\label{crit-curve1}
\ee
At this point
the critical degrees of freedom are represented by 
a degenerate doublet of massless Majorana fermions with a marginal
current-current interaction $g_{\parallel}(h_c)$. This is a typical
Gaussian [U(1)] criticality with central charge $C = 1$.
The vicinity of the critical point where the Majorana doublet
($\xi^1, \xi^2$) becomes very soft is 
described by the off-critical Askin-Teller model, or equivalently, the
quantum sine-Gordon model (SGM) for a scalar field $\Phi$:
\be
{\cal H}_{\rm t;eff} \simeq 
 \frac{v}{2}[(\p_x \Phi)^2 + (\p_x \Theta)^2]
- \frac{m_d}{\pi\alpha} \cos \sqrt{4\pi K} \Phi.
\label{sg}
\ee
Here $\Theta$ is the field dual to $\Phi$,
the parameter
\be
K = 1 - (g_{\parallel}/2\pi v) + O(g^2 _{\parallel})
\label{K}
\ee
determines the (coupling-dependent) compactification radius of the field $\Phi$,
and
\[
m_d \propto m_t (h_c - h)/h_c.
\] 
In the vicinity of the critical point, the spectral gap scales as
the renormalized
mass of the SGM (\ref{sg}):
\be
M_d \propto \frac{v}{\alpha}
 \left(\frac{|m_d| \alpha}{v} \right)^{\frac{1}{2 - K}} {\rm sign} (m_d).
\label{phys.mass}
\ee

In section 5, we show [see Eq.(\ref{eff-final})] that, in the strong-coupling limit 
($J_{\perp}, h \gg J$),
the effective model describing the low-energy properties of the spin ladder
at $h \sim h_c$ represents an anisotropic (XXZ) spin-1/2 Heisenberg chain with the 
parameter $\Delta$ close to 1/2 and an effective staggered magnetic field
$h^* \sim h - h_c$. In the continuum limit, this quantum lattice model
transforms to the SGM (\ref{sg}). The only difference between
the weak-coupling and strong-coupling regimes is that
in the latter case the parameter $K$ is close to 3/4.

The description of the low-energy part
of the spectrum of the original model (\ref{total-ham-dens}) in terms of the effective
anisotropic
spin-1 chain (\ref{eff.trip.ham}) holds if $h \ll J_{\perp}$. As follows from
(\ref{crit-curve1}), this condition is satisfied 
at $h<h_c$ and also
in some region above the critical field.
However, if the field reaches values $h \sim |m_s|$,
the singlet mode becomes as important as the triplet
ones, and 
the effective model (\ref{eff.trip.ham}) is no longer applicable. 
This regime is difficult to tackle analytically. On the other hand, if the field
$h$ is further increased and occurs in the range $|m_s| \ll h \ll J$,
the role of the interchain
exchange $J_{\perp}$ becomes subdominant, and the original model reduces to two decoupled 
Heisenberg chains in a weak staggered
magnetic field. In this case exact results are available, because in the scaling limit 
each such chain is described by a SGM with
a coupling constant $\beta^2 = 2\pi$ (see e.g. Ref.\cite{gnt}, chapter 22):
\bea
&&{\cal H} \to \sum_{j=1,2} {\cal H}_j,\nn\\
&& {\cal H}_j = \frac{v}{2} \left[(\p_x \Theta_j)^2 + (\p_x \Phi_j)^2 \right]
- \frac{\lambda(\bar{h})}{2\pi\alpha} \sin \sqrt{2\pi} \Phi_j,
\label{sg-2pi}
\eea
where $\lambda(\bar{h}) \sim \bar{h}$ (see \cite{essler2,oshikawa2} for
an accurate estimation of $\lambda(\bar{h})$).
(Technically, this mapping can be achieved either directly, i.e. 
using the rules of Abelian bosonization of the S=1/2 Heisenberg chain\cite{lp,ah},
or by establishing the correspondence between
two pairs of the Majorana fields and two bosonic fields, 
\[
(\xi_1, \xi_2) \leftrightarrow \Phi_+, ~(\xi_3, \xi_4) \leftrightarrow \Phi_-,~~~
\Phi_{\pm}= \frac{\Phi_1 \pm \Phi_2}{\sqrt{2}},
\]
and using formulas (\ref{br3}) of Appendix A
to bosonize the magnetic term $\s_1\s_2\mu_3\mu_4$.)

\subsection{Projecting operators onto the low-energy sector}

Since the fourth (singlet) Ising system has the largest
energy gap and stays ordered across the transition ($m_s < 0$),
at energies $\omega \ll |m_s|$ the order operator $\s_4$ can be replaced
by its nonzero expectation value $\la \s_4 \ra \sim (\alpha/l_s)^{1/8}$.
Under this substitution
the relative staggered magnetization ${\bf n}^-$ and dimerization field
$\eps^-$, defined in (\ref{n-}), (\ref{eps-pm}) become projected onto
the low-energy, triplet sector of the model described by the effective
action (\ref{eff.trip.action}), (\ref{eff.trip.ham}):
\bea
&&{\bf n}^- \to (\alpha/l_s)^{1/8} {\bf N}^- \nn\\
&& {\bf N}^- \sim (1/\alpha)
\left( \s_1\mu_2\mu_3, ~\mu_1 \s_2 \mu_3, ~\mu_1 \mu_2 \s_3 \right)
\label{n-:proj} \\
&&\eps^- \to (\alpha/l_s)^{1/8} E^-,
~~ E^-\sim (1/\alpha) \s_1 \s_2  \s_3 
\label{eps-:proj} 
\eea
On the other hand, the total staggered magnetization ${\bf n}^+$ and dimerization field
$\eps^+$ are both proportional to the disorder operator $\mu_4$
 whose correlations are exponentially decaying at
short distances, $r \sim l_s$
(see Eq.(\ref{mu4-mu4-corr})).
 Therefore one might conclude that these fields are short-ranged,
and the spectral weight of their fluctuations is only nonzero in
the high-energy region $\omega \sim |m_s|$. 
By the same argument,
the smooth part of the relative magnetization, ${\bf K}$, Eq.(\ref{vec-currents-maj}),
proportional to the singlet Majorana field $\xi^4$, would also
appear as a ``high-energy'' field.
However, this conclusion cannot be correct for the following reason.

It is true that, once the high-energy singlet modes are integrated out, 
the operator ${\bf n}^+$ defined in (\ref{n+})
has no projection onto the lower-energy sector. 
However, (\ref{n+}) is the {\sl zeroth-order}
definition of this operator with respect to the staggered field
which couples the high- and low-energy modes.
In fact, apart of the always existing short-ranged part,
the operator ${\bf n}^+$ contains a 
strongly fluctuating piece, which originates from a
finite admixture
of low-energy states occuring already in the first order in $h$. 
This can be easily understood from the fact that
reduction of the
original action of the model to the effective one
is equivalent to a unitary transformation of the quantum Hamiltonian of the system
that projects it to the subspace of the low-energy states. But the same unitary
transformation should be applied to physical operators to single out their low-energy
projections.

The low-energy projections of seemingly high-energy operators
can be extracted from the second-order perturbative corrections to the corresponding
correlation functions. Equivalently (and more formally), 
this can be done by fusing a local operator ${\cal O}_0({\bf r})$, 
originally defined as a short-ranged field 
with the perturbative part of the total action,
\bea
&&{\cal O}_0({\bf r}) \to {\cal O}({\bf r}) = {\cal O}_0 ({\bf r}) e^{-S_{st}} \nn\\
&&~~= {\cal O}_0({\bf r}) + \left(\frac{h}{v}\right)
\int \rd^2 {\bf r}_1  \la {\cal O}_0 ({\bf r})  n^+ _z ({\bf r}_1)\ra_s + O(h^2),
\label{op-transmut}
\eea
and averaging the first-order term over the fast singlet modes. 
This term is just the low-energy projection we are looking for
(it can be easily checked that the marginal part of the perturbation,
given by the operator ${\cal O}_{g}$ in (\ref{pertur-action}), yields no mapping
onto the low-energy triplet sector).
This is essentially an
``integrating-out'' procedure but this time applied to the correlation functions
rather than the action itself.

In Appendix \ref{admix} we show that the projection of the total staggered magnetization
onto the whole triplet sector is of the form:
\bea
&&n^+ _z \to \la n^+ _z \ra \nn\\
&&+~ \ri A_n \left(\frac{h}{|m_s|}\right) \left(\frac{l_s}{\alpha}\right) 
(\xi^1 _{R}\xi^1 _{L} + \xi^2 _{R}\xi^2 _{L}
- \xi^3 _{R}\xi^3 _{L}) + \cdots, \label{proj-n+z} \\
&&n^+ _x \to \ri A_n \left(\frac{h}{|m_s|}\right) \left(\frac{l_s}{\alpha}\right)
\left(\xi^1 _{R} \xi^3 _{L} + \xi^3 _{R} \xi^1_{L} \right)
+ \cdots , \label{proj-n+x}\\
&&n^+ _y \to \ri A_n \left(\frac{h}{|m_s|}\right) \left(\frac{l_s}{\alpha}\right)
\left(\xi^2 _{R} \xi^3 _{L} + \xi^3 _{R} \xi^2 _{L} \right)
+ \cdots ,
\label{proj-n+y}
\eea
where $\la n^+ _z \ra$ is the average staggered magnetization induced by the field
(see section V), 
$A_n$ is a numerical constant and the dots stand for
the high-energy parts of the operators. In that Appendix we also derive
the first-order low-energy projection of the relative smooth magnetization
of the ladder:
\bea
&&K_x  \to - \ri A_K
 \left(\frac{h}{|m_s|}\right)\left( \frac{\alpha}{l_s} \right)^{1/8}
\left(\frac{l_s}{v}\right) \p_{\tau} N^- _y , \label{proj-Kx}\\
&&K_y  \to  - \ri A_K
 \left(\frac{h}{|m_s|}\right)\left( \frac{\alpha}{l_s} \right)^{1/8}
\left(\frac{l_s}{v} \right) \p_{\tau} N^- _x ,\label{proj-Ky}\\
&&K_z  \to  A_K
 \left(\frac{h}{|m_s|}\right)\left( \frac{\alpha}{l_s} \right)^{1/8}
l_s \p_{x} E^- ,\label{proj-Kz}
\eea
where $A_K$ is another numerical constant and the fields
${\bf N}^-$ and $E^-$ are defined in (\ref{n-:proj}), (\ref{eps-:proj}).

\section{Dynamical structure factor in the presence of a staggered field}

The scattering cross section measured in neutron scattering experiments
is proportional to the dynamical structure factor $S^{\alpha\beta} (\omega, Q)$.
In this section we discuss some general properties of $S^{\alpha\beta}
(\omega, Q)$ in the case where a quantum spin chain or ladder
is subject to parity-breaking external perturbations, such as a
staggered magnetic field or explicit dimerization. 

Consider first the case of a single Heisenberg chain.
The dynamical structure factor is defined as the Fourier transform of the
spin-spin correlation function
\bea
&&S^{\alpha\beta} (\omega, Q) \nn\\
&&= \frac{1}{2\pi N}\sum_{n,m=1}^{N} \int_{-\infty}^{\infty}
\rd t e^{\ri \omega t - \ri Q (n-m)} 
\la S^{\alpha}_n (t)  S^{\beta}_m (0)\ra. \label{Str-Fac:def}
\eea
Here we have set the lattice spacing $a_0 = 1$.
For a translationally invariant, antiferromagnetic spin chain
the spin-spin correlation function has the following asymptotic structure
\bea
&&\la S^{\alpha}_n (t)  S^{\beta}_m (0)\ra\nn\\
&& = F_1^{\alpha\beta}(t,n-m)
+(-1)^{n-m}F_2^{\alpha\beta}(t,n-m), \label{corr-fun-tr.inv}
\eea
where $F_{1}(t,n)$ and $F_{2}(t,n)$ are slowly varying functions of $n$ and $t$.
According to (\ref{spin-decom}), in the continuum limit these reduce to
the correlation functions of the smooth and staggered magnetization, 
$\la J^{\alpha} (t,x) J^{\beta}(0,0) \ra$ and 
$\la n^{\alpha} (t,x)  n^{\beta} (0,0) \ra$, and thus  determine the dynamical
structure factor in the vicinity
of two {\sl different} points: $Q \approx 0$ and $Q \approx \pi$.

When a staggered magnetic field is applied to a spin chain, the situation may seem 
to be different. Indeed,
the translational symmetry of the underlying lattice is broken and
the period of the induced magnetic structure is $2$. 
The spin excitation spectrum is now defined in the reduced Brillouin zone
$-\pi/2 < q \leq \pi/2$, with the points $q=0$ and $q=\pi$
identified. At the same time, due to the lowered translational symmetry,
the asymptotical expression (\ref{corr-fun-tr.inv}) for
the spin-spin correlation function will contain extra oscillating pieces
\bea
&&\langle S^\alpha_n(t) S^\beta_m(0)\rangle=
F_1^{\alpha\beta}(t,n-m)
+(-1)^{n-m}F_2^{\alpha\beta}(t,n-m)\nn\\
&&\qquad +(-1)^{n}F_3^{\alpha\beta}(t,n-m)
+(-1)^{m}F_4^{\alpha\beta}(t,n-m),
\label{f1to4}
\eea
where $F_3 (t,n)$ and $F_4 (t,n)$ are smooth functions that
transform in the continuum limit to the mixed correlators
$\la n^{\alpha}(t,x) J^{\beta}(0,0) \ra$ and $\la  J^{\alpha}(t,x)
n^{\beta}(0,0) \ra$ respectively. These correlators are nonzero in the
presence of the staggered field, and the question is whether they
contribute to the dynamical structure factor.

The answer to this question is negative. At a formal level, this can
be shown as follows. Note that due to the broken one-site translational
symmetry the spin-spin correlation function should be expanded in a double
Fourier series
\be
\langle S^\alpha_n(t) S^\beta_m(0)\rangle=
\frac{1}{N} \sum_{k,k'} e^{\ri (kn-k'm)} \tilde{F}^{\alpha\beta}_{k,k'}(t),
\label{doub-Four}
\ee
where $k$ and $k'$ vary within the paramagnetic Brillouin zone 
($-\pi < k \leq \pi$), and the double-periodicity
requires that $\tilde{F}^{\alpha\beta}_{k,k'} \neq 0$ 
if $k=k'$ or $k=k' +\pi$.
Mapping onto the reduced Brillouin zone yields:
\bea
\langle S^\alpha_n(t) S^\beta_m(0)\rangle
&&= \frac{1}{N} \sum_{|q|<\pi/2} e^{\ri q(n-m)}\nn\\
&&\times
[\tilde{F}^{\alpha\beta}_{q,q}(t)
+ (-1)^{n-m} \tilde{F}^{\alpha\beta}_{q + \pi,q+ \pi}(t) \nn\\
&& +~ (-1)^{n} \tilde{F}^{\alpha\beta}_{q + \pi,q}
+ (-1)^{m} \tilde{F}^{\alpha\beta}_{q,q+ \pi}(t)]
\label{doub-fourier-ser}
\eea
At small $q$, $\tilde{F}_{q,q}$, $\tilde{F}_{q + \pi,q+ \pi}$,
$\tilde{F}_{q + \pi,q}$ and $\tilde{F}_{q,q+ \pi}$ are the Fourier transforms
of the smooth functions $F_1 (n)$, $F_2 (n)$, $F_3 (n)$ and $F_4 (n)$ respectively. 
Substituting (\ref{doub-Four}) into (\ref{Str-Fac:def}) we find that
\bea
S^{\alpha\beta} (\omega, Q) =  \frac{1}{2\pi}\sum_{l,l'=-\infty}^\infty
\tilde{F}^{\alpha\beta}_{Q+2\pi l, Q+2\pi l'}(\omega). \label{str-factor-formal}
\eea
From (\ref{str-factor-formal}) it follows that the structure factor is 
$2\pi$-periodic in $Q$. Secondly, the r.h.s. of (\ref{str-factor-formal})
does not contain off-diagonal
matrix elements, $\tilde{F}_{Q,Q+\pi}$ and $\tilde{F}_{Q+\pi,Q}$, implying
that mixed correlators do not contribute to $S^{\alpha\beta} (\omega, Q)$.
Therefore
\begin{equation}
S^{\alpha\beta}(\omega,Q)=\frac{1}{2\pi}
\cases{
{\tilde{F}^{\alpha\beta} _1(\omega,Q)} & {\rm if\ $Q\approx 0$}\cr
{\tilde{F}^{\alpha\beta} _2(\omega,Q-\pi)} & {\rm if\ $Q\approx \pi$}\ ,\cr
}
\label{salbe}
\end{equation}
where at small $Q$ and $\omega$
$\tilde{F}^{\alpha\beta} _1(\omega,Q)$ and $\tilde{F}^{\alpha\beta} _2(\omega,Q)$ 
are Fourier transforms of the correlation functions \hfill\break
$\la  J^{\alpha}(t,x) J^{\beta}(0,0) \ra$ and $\la  n^{\alpha}(t,x) n^{\beta}(0,0) \ra$, 
respectively.

The same conclusion can be reached within an equivalent but
somewhat more appealing picture of
diatomic cells. Define a magnetic unit cell made of two sites,
$(2n-1,2n)$, and denote the corresponding
spin operators by $S^\alpha_{2l} =\S_l$ and $S^\alpha_{2l-1}=\T_l$.
There altogether are four different spin-spin correlation functions
\bea
g^{\alpha\beta}_1(t,l-l')&=&\langle {\cal S}^{\alpha} _l(t)\ {\cal S}^{\beta} _{l'}(0)\rangle\ ,\nn\\
g^{\alpha\beta}_2(t,l-l')&=& \langle {\cal T}^{\alpha} _l(t)\ {\cal T}^{\beta} _{l'}(0)\rangle\,\nn\\
g^{\alpha\beta}_3(t,l-l')&=&\langle {\cal S}^{\alpha} _l (t)\ {\cal T}^{\beta} _{l'}(0)\rangle\ ,\nn\\
g^{\alpha\beta}_4(t,l-l')&=&\langle {\cal T}^{\alpha} _l(t)\ {\cal S}^{\beta} _{l'} (0)\rangle\ ,
\eea
whose Fourier transforms 
\bea
\tilde{g}_a(\omega,q)&=&\sum_{l=1}^{N/2}\int_{-\infty}^\infty \rd t
\exp(\ri \omega t-\ri q [2la_0])\ g_a(t,l)
\eea
have the periodicity of the reduced Brillouin zone
\bea
\tilde{g}_a(\omega,q+\pi)=\tilde{g}_a(\omega,q)\ .
\eea
It then follows from the definition (\ref{Str-Fac:def}) that
\bea
S^{\alpha\beta}(\omega,Q)&=&
\frac{1}{4\pi} [
\tilde{g^{\alpha\beta}}_1(\omega,q) + \tilde{g}^{\alpha\beta}_2(\omega,q)\nn\\
&&+e^{-iQa_0}\tilde{g}^{\alpha\beta}_3(\omega,q)
+e^{iQa_0}\tilde{g}^{\alpha\beta}_4(\omega,q) ].
\label{S-vs-g}
\eea
This expression shows that,
although the spin
correlation functions $\tilde{g}_a(\omega,q)$ have the periodicity of
the reduced Brillouin zone, the dynamical structure factor does not; it rather
retains the periodicity of the paramagnetic Brillouin zone. Thus,
for the dynamical structure factor $S^{\alpha\beta} (\omega,Q)$,
the points $Q=0$ and $Q=\pi/a_0$ are {\sl inequivalent}
even in the presence of a staggered magnetic field.

The functions $g_a (t,l)$ can be easily expressed in terms of the functions
$F_a (t,n)$, defined in (\ref{f1to4}). Using then (\ref{S-vs-g}) one finds that

\bea
S^{\alpha\beta}(\omega,Q)&=&\frac{1}{2\pi}
\int_{-\infty}^\infty \rd t\sum_{n=1}^N e^{\ri\omega t+\ri Qn}\ F^{\alpha\beta} _1(t,n)\nn\\
&&+\int_{-\infty}^\infty \rd t\sum_{n=1}^N e^{\ri \omega t+
\ri [Q+\pi]n }\ F^{\alpha\beta} _2(t,n) .
\eea
Using the fact that $F_a(t,n)$ are slowly varying functions
of $n$, we finally arrive at the result (\ref{salbe}) where
\bea
\tilde{F}_a^{\alpha\beta}(\omega,p)&=&
\int_{-\infty}^\infty dt\sum_{n=1}^{N}
\exp(i\omega t+ip n)\ F^{\alpha\beta} _a(t,n)\ .
\label{salbe2}
\eea

The generalization to the case of the two-leg ladder is
straightforward. The structure factor is defined by 
\begin{eqnarray}
&&S^{\alpha\beta}(\omega,q,q_\perp)= 
\frac{1}{4\pi N}
\sum_{a,b=1}^2\sum_{n,m=1}^N\int_{-\infty}^\infty \rd t\ e^{\ri \omega t}\nn\\
&&\quad\times\ e^{-\ri q(n-m)-iq_\perp(a-b)}\
\langle S^\alpha_{a,n}(t),S^\beta_{b,m}(0)\rangle ,
\label{DSF}
\end{eqnarray}
where we have introduced a transverse momentum $q_\perp$ that can take
only the two values, $0$ and $\pi$. Information about the low-energy
part of the spin fluctuation spectrum is contained in the staggered
($q\approx \pi$) and smooth ($q\approx 0$) parts of the structure
factor for both values of $q_{\perp}$. All these four cases will be
considered separately. Below we adopt the notation
\[
s^2 = \omega^2 - q^2 v^2,
\]
where $q$ stands for a small momentum deviation either from
0 or $\pi$.
\section{Dynamical structure factor in the weak-coupling limit}
\label{sec:secIV}
In this section we determine the dynamical structure factor at low
energies in the weak-couling regime. In subsections IV A -- D it
will be assumed that the staggered magnetic field is much smaller
than the singlet gap $|m_s|$. 
In this case the relevant correlation functions can be estimated
using the effective action in the triplet sector and the low-energy
projections of the corresponding physical fields, discussed in the
preceeding section. In subsection IV E we will consider another limiting
case, $|m_s| \ll h \ll J$, which will be treated on the basis of
the model (\ref{sg-2pi}).

\subsection{Structure factor at q$_{\vperp}=\vpi$, q $\approx$ $\vpi$}

Of primary importance is the evolution of the coherent triplet peak displayed
by the dynamical structure factor of the Heisenberg ladder
under the action of the gradually increasing staggered magnetic
field. This information is contained in the spectral properties of the
relative staggered magnetization ${\bf n}^-$
whose low-energy projection is given by (\ref{n-:proj}).
We therefore start our discussion
with the case $q_{\perp} = \pi$ and consider the real-space--imaginary-time
correlation functions
\be
D^{(-)}_{\alpha\alpha} ({\bf r}) = (\alpha/l_s)^{1/4}\la N^{-}_{\alpha} ({\bf r}) 
N^- _{\alpha} ({\bf 0}) \ra ~~(\alpha=x,y,z). 
\label{D-definition}
\ee
The corresponding structure factor is obtained by Fourier
transformation, $D^{(-)}_{\alpha\alpha} ({\bf r}) \to
D^{(-)}_{\alpha\alpha} ({\bf q})$ (${\bf q} = (q, \omega_n/v)$), 
and subsequent analytical continuation to the upper complex
$\omega$-plane ($\ri \omega_n \to \omega + \ri \delta$).

\subsubsection{$h<h_c$}

For fields smaller than $h_c$ the leading asymptotics of
$D^{(-)}_{\alpha\alpha} ({\bf r})$ are given by 
\bea
&&D^{(-)}_{xx} ({\bf r}) = D_{yy} ({\bf r})
\propto (1/\alpha^2) (\alpha^2/l_s l_3)^{1/4}
\nn\\
&&\qquad\qquad\qquad\qquad \times
\la \s_1 ({\bf r}) \mu_2 ({\bf r}) \s_1 ({\bf 0}) \mu_2 ({\bf 0})  \ra,
\label{D-transv} \\
&& D^{(-)}_{zz} ({\bf r})  
\propto (1/\alpha^2) (\alpha/l_s)^{1/4}\nn\\
&& \qquad\qquad \times\la  \mu_1 ({\bf r})\mu_2 ({\bf r})\mu_1 ({\bf 0})\mu_2 ({\bf 0})\ra~
\la  \s_3 ({\bf r}) \s_3 ({\bf 0})\ra,
\label{D-parallel}
\eea
where $l_3 \sim v/m_3$. Here and in what follows it is
assumed that the third Ising system ($\xi^3$) decouples from the 
doublet $(\xi^1, \xi^2)$. This assumption is certainly correct in the
vicinity of the critical point and holds on a qualitative level
everywhere in the massive phases at $h \neq h_c$.

Using the local bosonic representation of the product $\s_1 \mu_2$, given
by Eq.(\ref{br4}) of Appendix A, 
and properly rescaling the dual field $\Theta$, $\Theta \to
(1/\sqrt{K}) \Theta$, we find that the correlator
in the r.h.s. of (\ref{D-transv}) reduces to that of vertex operators of
the dual field in the SGM (\ref{sg}) which, in turn,
can be estimated by means of the formfactor bootstrap approach
\cite{karo,smirnov,FF,lukPLA,lz2001} 
\be
\la \cos \sqrt{\frac{\pi}{K}} \Theta ({\bf r}) \cos \sqrt{\frac{\pi}{K}} \Theta ({\bf 0})
\ra \sim Z(K) \frac{e^{-M_d r/v}}{\sqrt{M_d r/v}}.
\label{corr.dual.exp}
\ee
Inspecting the r.h.s. of (\ref{D-parallel}), we notice that
the disorder parameters $\mu_1$ and $\mu_2$ have nonzero expectation values
($m_d > 0$). On the other hand,
since $m_3 > 0$, the correlator $\la \s_3 ({\bf r}) \s_3 ({\bf 0}) \ra$
has the asymptotic form (\ref{mu4-mu4-corr}) with $l_s$ replaced by $l_3$.
The obtained asymptotics for the correlation functions $D^{(-)}_{xx} ({\bf r})$ 
and $D^{(-)}_{zz} ({\bf r})$ lead to the following expressions for
the transverse and longitudinal
components of the dynamical structure factor:
\bea
S^{+-} (\omega,\pi +q,\pi) &=& C_{\perp} (h)
\delta \left(s^2  -  M^2 _d \right),\label{xx:below}\\
S^{zz} (\omega,\pi +q,\pi) &=& C_{\parallel} (h)
\delta \left(s^2  - m^2 _3 \right),\label{zz:below}
\eea
where 
\be
C_{\perp}(h), C_{\parallel}(h) \sim (|m_s| m_3 M^2 _d)^{1/4}.
\label{C-perp;C-paral}
\ee
We note that the incoherent continua that contribute to the dynamical
susceptibilities at higher energy can be calculated by using the
results of Ref.\cite{wu} (see e.g. \cite{spin1} for a similar
calculation).

Thus, the coherent triplet magnon peak of the isotropic Heisenberg spin ladder,
originally located at 
the frequency $\omega = m_t$,
is split by the field in two peaks: the doublet ($S^z = \pm 1$) peak at $\omega =
M_d < m_t$  and the $S^z = 0$ peak at $\omega = m_3 > m_t$. Therefore, the phase
occuring at $h < h_c$ represents an anisotropic spin liquid with coherent
longitudinal and transverse magnon excitations having different, field
dependent mass gaps.

Upon increasing the field, the two peaks move in opposite
directions. When the critical field is approached, the background of
multiparticle states with thresholds $3M_d$, $5 M_d$, ..., and the
doublet peak merge, and, at criticality, the 4-point Ising corelation
functions in (\ref{D-transv}) and (\ref{D-parallel}) follow a
power-law behavior.

\subsubsection{$h=h_c$}

As follows from the bosonic representation of the products $\s_1 \mu_2$ and $\mu_1 \mu_2$
(see Eqs.(\ref{br3}),(\ref{br4}) of Appendix A), at $h=h_c$ the four-point Ising correlators
in (\ref{D-transv}) and (\ref{D-parallel}) transform to those of vertex operators in
a Gaussian model and therefore display a power-law decay. One easily finds that
\bea
&& D^{(-)}_{xx} ({\bf r}) \propto \left( \frac{\alpha}{r} \right)^{1/2K},\nn\\
&& D^{(-)}_{zz} ({\bf r}) \propto \left( \frac{\alpha}{r} \right)^{K/2} 
\frac{e^{- r/l_3}}{\sqrt{r/l_3}}.
\label{corr-crit}
\eea
After the Fourier transformation and analytic continuation we obtain:
\bea
&&S^{+-} (\omega,\pi+q,\pi) = C' _{\perp}
 \left( \frac{v^2}{\alpha^2 s^2} \right)^{1 - \frac{1}{4K}}\theta(s^2),
\label{trans-crit} \\
&&S^{zz} (\omega,\pi+q,\pi)\nn\\
&& \qquad = C' _{\parallel}
 \left[ \frac{v^2}{\alpha^2 (s^2 - m^2 _3)} \right]^{1-\frac{K}{2}}
\theta(s^2 - m^2 _3),
\label{paral-crit}
\eea
where 
\bea
&&C' _{\perp} \sim \left(\frac{\alpha}{v}\right)
\left(\frac{\alpha^2}{l_s l_3}\right)^{\frac{1}{4}}, \nn\\
&& C' _{\parallel} \sim \left(\frac{\alpha}{v}\right)
\left(\frac{\alpha}{l_s}\right)^{1/4} 
\left(\frac{\alpha}{l_3}\right)^{\frac{1}{4} - \frac{K}{2}}
\label{C'-perp;C'-paral}
\eea
and
$\theta(x)$ is the Heaviside step function. Both the transverse
an longitudinal susceptibilities feature incoherent scattering
continua with thresholds at zero energy and $\omega=m_3$ respectively.

\subsubsection{$h_c < h \ll |m_s|$}

At $h > h_c$ the mass $m_d$ becomes negative and the doublet of Ising
systems occurs in the ordered phase, whereas the third Ising system stays
disordered ($m_3 > 0$). 
The behavior of $D^{(-)}_{xx} ({\bf r})$ is just the same as at $h < h_c$ since
the duality transformation associated with the sign reversal of $m_d$,
i.e. $\mu_{1,2} \leftrightarrow \s_{1,2}$, 
keeps the correlator in the r.h.s. of (\ref{D-transv}) unchanged.
Therefore the coherent $|S^z| = 1$ magnon peak, which exists at $h < h_c$ and
disappears at the critical point, is recovered
in the $h>h_c$ phase. In contrast to this, the asymptotics of $D^{(-)}_{zz}
({\bf r})$ is changed, and at $h>h_c$ we find
\bea
D^{(-)}_{zz}({\bf r}) &\propto& 
\frac{1}{v\alpha^2} \left(|m_s|m_3M_d^2\right)^{1/4}
\frac{e^{- 2 |M_d| r/v}}{|M_d| r/v} 
\frac{e^{- m_3 r/v}}{\sqrt{m_3 r/v}}.\nn\\
\label{zz-corr:above.crit}
\eea
As a consequence, the Haldane spin liquid loses a part of its coherent
spectral weight at $q \sim \pi$:  
the $S^z = 0$ magnon is no longer seen in the longitudinal staggered structure factor
$S^{zz} (\omega, \pi+q, \pi)$
and is
replaced by an incoherent continuum of states with a threshold at
$\omega = 2 |M_d| + m_3$:
\be
S^{zz}(\omega,\pi+q,\pi) =  C''_{\parallel}
\theta[s^2 - (2|M_d| + m_3)^2].\\
\label{chi-par-above}
\ee
where
\be
C''_{\parallel}(h) \sim \left( \frac{\alpha}{v} \right)
\left( \frac{\alpha^2}{l_s l_3} \right)^{1/4}
\left(\frac{v}{(2|M_d| + m_3)\alpha} \right)^{\frac{1}{2}}.
\label{C''-paral}
\ee
We will show below that, in fact, the coherent $S^z = 0$ mode still exists at $h>h_c$ 
but its spectral weight is shifted towards small momenta ($q\sim 0$)
and is very small.

\subsection{Structure factor at q$_{\vperp}=\vpi$, q $\approx$ 0}

The structure factor $S^{\alpha\alpha}(\omega,q,\pi)$ is determined by
correlations of the relative magnetization
\bea
&&{\bf S}_{1,n} - {\bf S}_{2,n} \to {\bf K}(x),\nn\\
&& K^a = \ri (\xi^a_{R} \xi^4 _{R} +  \xi^a_{L} \xi^4 _{L}),
~a=1,2,3.
\label{relmag-vs-K}
\eea
The projections of the fields $K^a$ onto the low-energy sector
are given by Eqs (\ref{proj-Kx})--(\ref{proj-Kz}).

Consider first the longitudinal structure factor.
Since the third Ising system remains disordered across the transition ($m_3 > 0$),
$S^{zz}(\omega, q, \pi)$ will be nonzero only at frequencies $\omega
\geq m_3$. This follows from (\ref{proj-Kz}) and
(\ref{eps-:proj}). Note that the operator $E^-$ in (\ref{eps-:proj}) 
is related to $N^- _z$, Eq.(\ref{n-:proj}), by a duality
transformation in the doublet sector. Therefore the large-distance
asymptotics of $\la E^- ({\bf r}) E^- ({\bf 0})\ra$
can be obtained from those of $D^{(-)}_{zz} ({\bf r})$, 
estimated in the preceeding subsection,
by interchanging the cases $h<h_c$ and $h>h_c$.
This leads to the following results:

\noindent
\underline{$h<h_c$:}
\bea
&&S^{zz}(\omega, q, \pi) \nn\\
&&= C''_{\parallel}(h)
 \left(\frac{h}{m_s} \right)^2
\left( q l_s\right)^2 \theta[s^2  - (2M_d + m_3)^2];
\label{long:h<hc}
\eea

\noindent
\underline{$h=h_c$:}
\bea
&&S^{zz}(\omega, q, \pi)=\nn\\
&&C' _{\parallel}
 \left(\frac{h}{m_s} \right)^2 \left( q l_s\right)^2
\left[ \frac{v^2}{\alpha^2 (s^2 - m^2 _3)} \right]^{1-\frac{K}{2}}
\theta(s^2 - m^2 _3);
\label{long:h=hc}
\eea

\noindent
\underline{$h>h_c$:}
\be
S^{zz}(\omega, q, \pi) = C _{\parallel}(h)
 \left(\frac{h}{m_s} \right)^2 \left( q l_s\right)^2
\delta (s^2 - m^2 _3);
\label{long:h>hc}
\ee
where the prefactors $C _{\parallel}$,
$C' _{\parallel}$, $C''_{\parallel}$ are given by (\ref{C-perp;C-paral}),
(\ref{C'-perp;C'-paral}),(\ref{C''-paral}).
We see that at $h>h_c$ the coherent $S^z = 0$ magnon mode is seen in the
small-$q$ part of the dynamical structure factor. However, its spectral
weight is proportional to $\left( q l_s\right)^2$ and thus small.

Consider now the transverse structure factor. Using the definitions
(\ref{proj-Kx}), (\ref{proj-Ky}) and the known expression
for the structure factor $S^{\pm}(\omega, \pi + q, \pi)$ obtained
in section IV A, we find that

\noindent
\underline{$h\neq h_c$:}
\bea
&&S^{\pm}(\omega, q, \pi)\nn\\
&& = C_{\perp} (h) \left( \frac{h}{m_s} \right)^2 \left( \frac{\omega}{m_s} \right)^2
\delta(s^2 - M^2 _d); \label{trans:h-neq-0}
\eea

\noindent
\underline{$h = h_c$:}
\bea
&&S^{\pm}(\omega, q, \pi)\nn\\
&& = C' _{\perp} \left( \frac{h}{m_s} \right)^2 \left( \frac{\omega}{m_s} \right)^2
\left( \frac{v^2}{s^2 \alpha^2} \right)^{1-\frac{1}{4K}} \theta(s^2);
\label{trans:h=0}
\eea
where $C_{\perp}$ and $C' _{\perp}$ are given by (\ref{C-perp;C-paral}),
(\ref{C'-perp;C'-paral}).
Comparing the results (\ref{xx:below}) and (\ref{trans:h-neq-0}), we see that
at any nonzero $h$ (except for the transition point),
the coherent transverse ($|S^z| =1$) magnon mode is seen both in the
staggered ($q\approx\pi$) {\sl and} smooth ($q\approx 0$) parts of the structure
factor $S^{\pm} (\omega, q, \pi)$, with the ratio of the
spectral weights at
$q\approx 0$ and $q\approx\pi$ being of the order of $(h/m_s)^2 (M_d/m_s)^2$.


\subsection{Structure factor at q$_{\vperp}$=0, q $\approx$ $\vpi$}


The structure factor $S^{\alpha\beta}(\omega, \pi + q, 0)$ is determined
by dynamical correlations of the total staggered magnetization,
${\bf n}^+$, whose low-energy projection is given by expressions
(\ref{proj-n+z})--(\ref{proj-n+y}).

\subsubsection{Longitudinal structure factor}

As follows from (\ref{proj-n+z}),
$S^{zz}(\omega, \pi + q, 0)$ displays a broad continuum of states with
the lowest-energy threshold equal to $2|M_d|$. 
If $|h-h_c| \ll h_c$, at frequencies much less than $m_3$ only the doublet modes are to be taken
into account. In this case we have
\bea
&&D^{(+)}_{zz} ({\bf r}) = A^2 _n \left( \frac{h}{m_s} \right)^2
\left( \frac{l_s}{\alpha} \right)^2 K_d ({\bf r}), \label{D-vs-K_d}\\
&& K_d ({\bf r}) =
\la \vare_d ({\bf r}) \vare_d ({\bf 0}) \ra, \label{K_d-def} 
\eea
where 
$
\vare_d = \ri  \left(\xi^1 _R \xi^1 _L + \xi^2 _R \xi^2 _L \right)   
$
is the energy-density operator in the doublet sector.
A simple calculation, based on the assumption that
the doublet fermions are free, leads to the result (see Appendix D):
\be
\Im m K_d (q,\omega + \ri \delta) = \frac{1}{2v}\frac{\sqrt{s^2 - 4m^2 _d}}{s}. 
\label{ed-ed-imaginary1}
\ee

In fact, the square-root behaviour of the structure factor near the threshold
is {\sl universal}: the effect of the interaction between the Majorana fermions
$\xi^1$ and $\xi^2$ shows up only in the mass and velocity renormalization
and the interaction dependent prefactor $Z_d (K)$:
\bea
&&S^{zz}(\omega, \pi + q, 0) \propto \frac{1}{2v}Z^2 _d (K)\left( \frac{h}{|m_s|} \right)^2
\left( \frac{l_s}{\alpha} \right)^2 \nn\\
&&~~~~\times
\frac{\sqrt{s^2 - 4 M^2 _d}}{s}~
\theta(s^2 - 4 M^2 _d)\ .
\label{Szz-stag-0}
\eea 
This is confirmed by the calculation done in section VI E,
which proceeds from the bosonized correlator (\ref{K_d-def}) 
\be
K_d ({\bf r})
\propto \la \cos \sqrt{4\pi K} \Phi({\bf r})  \cos \sqrt{4\pi K} \Phi({\bf 0})\ra
\label{ed-ed-bos}
\ee
and automatically takes into account interaction effects in the doublet
sector within the formfactor approach\cite{karo,smirnov,FF} 
to the SGM (\ref{sg}).
In that calculation, the crucial fact is that, in the
range $3/4 \leq K < 1$, the spectrum of the SGM (\ref{sg})
consists of massive quantum solitons ($s$) and antisolitons ($\bar{s}$) 
with the mass $M_d$, and one soliton-antisoliton bound state
(breather) with the mass $M_1$ given by Eq. (\ref{1breather}).
The latter, however, is odd under parity (charge) conjugation
(\ref{parity1}) and as a result the formfactor of the parity-symmetric
operator $\cos \sqrt{4\pi} \Phi$ between the vacuum and the breather
state vanishes. Therefore the main contribution to the structure factor 
is due to the $s\bar{s}$ scattering continuum with a threshold at
$2|M_d|$, where $M_d$ is the single-soliton mass. This explains 
the universality of the square-root behavior (\ref{Szz-stag-0})
of the structure factor $S^{zz}(\omega, \pi + q, 0)$ near the threshold.

The result (\ref{Szz-stag-0}) neglects the contributions of
multiparticle processes with thresholds at higher energies ($4|M_d|$,
$6|M_d|$, etc). As $h \to h_c$ (i.e. $M_d \to 0$),
such processes become as important as the  two-particle ones. Exactly
at criticality the Majorana doublet becomes massless, and the
interaction in the doublet sector can no longer be ignored.
In this case 
$
K_d (r)  
\simeq (\alpha /r)^{2K},
$ 
and
\be
S^{zz}(\omega, \pi + q, 0) 
\propto
\frac{C(K)}{v} 
\left(\frac{v^2}{\alpha^2 s^2} \right)^{1-K}\theta (s^2),
\label{S-zz-0-crit}
\ee
where $C(K) \sim [2^{1-2K}/\Gamma^2 (K)] (h/m_s)^2 (l_s/\alpha)^2$.

\subsubsection{Transverse structure factor}

\begin{figure}[ht]
\begin{center}
\epsfxsize=0.45\textwidth
\epsfbox{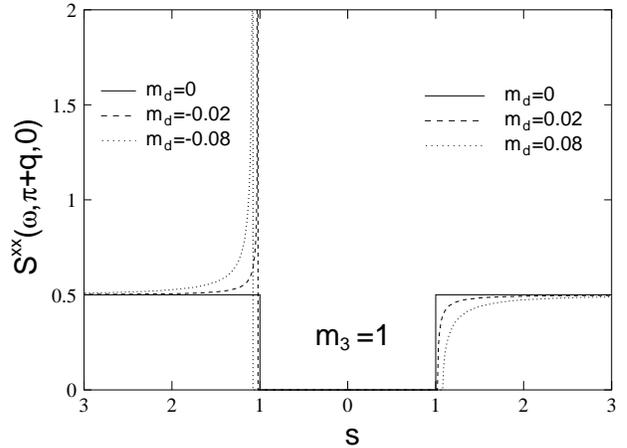 }
\end{center}
\caption{Transverse structure factor at $q \approx\pi, q_{\perp} = 0$. For more clarity,
the cases $h<h_c (m_d >0)$ and $h>h_c (m_d <0)$ are shown separately.} 
\label{fig:fig2}
\end{figure}

The transverse structure factor $S^{xx}(\omega, \pi+q, 0)$
can be estimated in a similar manner. The main difference is that now we have
two Majorana fermions ($\xi^1$ and $\xi^3$) with unequal masses ($m_d$
and $m_3$). Treating both fermions as free, simple calculations along
the lines of Appendix D give
\bea
S^{xx}(\omega, \pi + q, 0) \propto && \frac{\alpha}{v}\left( \frac{h}{|m_s|} \right)^2
\left( \frac{\alpha}{l_s} \right)^2 
\sqrt{\frac{s^2 - m^2 _+}{s^2 - m^2 _-}},\label{Sxx-stag-0}\\
&&
s^2 \geq {\rm max}~\{m^2 _+, m^2 _-\}, \nn
\eea 
where $m_{\pm} = m_3 \pm m_d$. For different signs of the doublet mass
$m_d \sim h_c - h$
the frequency dependence of the structure factor is qualitatively different
(see Fig.\ref{fig:fig2}).
At $h<h_c$ ($m_d > 0$) $S^{xx}(\omega, \pi + q, 0)$ follows a square-root increase
above the threshold $m_+$ which becomes steeper as $h \to h_c$. 
At $h > h_c$ $S^{xx}(\omega, \pi + q, 0)$ has a square-root singularity
at the threshold $m_-$ with an amplitude proportional to $|m_d|^{1/2}$.

Let us now discuss the role of the so far neglected interaction between the fermions.
As we have mentioned before, near the critical point the $S^z = 0$ collective modes
described by the field $\xi^3$ asymptotically decouple from the $S^z = \pm 1$ (doublet)
ones. Hence one can always assume that the Majorana fermion $\xi^3$
is free and massive. Concerning the fermion $\xi^2$ that belongs to
the interacting doublet sector, we may use the bosonized expressions
for the chiral components of $\xi^2$ (see
Eqs.(\ref{fer-bosL}),(\ref{fer-bosR})) and employ a formfactor
expansion in the model (\ref{sg}) to show that
\bea
\la \xi^2 _R (x,\tau) \xi^2 _R (0,0) \ra &=& Z(K) K_1 (|M_d z|) 
\left(\frac{z}{\bz} \right)^{1/2}, \nn\\
\la \xi^2 _L (x,\tau) \xi^2 _L (0,0) \ra &=& Z(K) K_1 (|M_d z|) 
\left(\frac{\bz}{z} \right)^{1/2}, \nn\\
\la \xi^2 _R (x,\tau) \xi^2 _L (0,0)\ra  &=& Z(K) K_0 (|M_d z|).
\label{maj-prop-LL}
\eea
Here $z=\tau + \ri x/v$, $\bz=\tau - \ri x/v$, and $K_n (x)$ are the MacDonald functions.
In (\ref{maj-prop-LL}) only the
one-particle form factor has been taken into account 
(the first correction involves a contribution of three-particle processes).
We see that the expressions (\ref{maj-prop-LL})
have the structure of the 2-point correlators of free massive Majorana fermions.
The information about the interactions in the doublet sector, i.e. the parameter
$K$ of the SGM, is contained
in the renormalized mass $M_d$  and the constant $Z(K)$.
This means that, up to this prefactor, the above result (\ref{Sxx-stag-0})
actually represent the first term of the exact
formfactor expansion and, thus, 
is universal close to the threshold as long as 
$h \neq h_c$.

According to (\ref{Sxx-stag-0}), at criticality the free-fermion approximation
leads to a step function $\theta(s^2 - m^2 _3)$.
However, this result must be revisited because at $M_d = 0$ the single-fermion
propagator $\la \xi^1 (x,\tau) \xi^1 (0,0) \ra$ transforms to that of a spinless
Tomonaga-Luttinger liquid. An estimation of $S^{xx}(\omega, \pi + q, 0)$, quite similar
to that done in Appendix \ref{app-transverse}, leads to a power-law behavior
\be
S^{xx} (\omega, q,0)\biggr|_{\rm crit}
= {\cal C}(\theta) 
\left( \frac{s^2 - m^2 _3}{m^2 _3} \right)^{2\theta},
\label{LL-threshold1}
\ee
where
\[
{\cal C}(\theta) \sim 
\frac{(\alpha/v)}{2^{1+2\theta} \Gamma(1+2\theta)}
\left(\frac{\alpha}{l_3}\right)^{2\theta}
\]
and
\be
2\theta = \frac{1}{2} \left( K + \frac{1}{K} \right) - 1
\label{2theta}
\ee
is
the critical exponent of the single-particle density of states in a
Tomonaga-Luttinger liquid.  We see that, due to an ``infrared
catastrophy'' caused by the interactions in the doublet sector,
the threshold discontinuity of $S^{xx} (\omega, \pi+q,0)$ 
transforms to a continuous dependence. However, due to the smallness
of $\theta$, the power-law increase of the transverse structure factor
is very steep.


\subsection{Structure factor at q$_{\vperp}$ = 0, q $\approx$ 0}


At small $q$ the structure factor $S^{\alpha\alpha} (\omega, q, 0)$
is determined by correlations of the smooth part of the
total magnetization density. In the continuum limit, the latter can be
expressed in terms of the triplet Majorana fields:
\bea
&&{\bf S}_{1,n} + {\bf S}_{2,n} \to {\bf I}(x),\nn\\
&& I^a = -(\ri/2)\eps^{abc}( \xi^b_{R}\xi^c_{R} + \xi^b_{L}\xi^c_{L} ).
\label{smooth-I}
\eea

\subsubsection{Longitudinal structure factor}

The total longitudinal current, $I^3 = -\ri \xi^1_{\nu}\xi^2_{\nu}$,
involves only the doublet Majorana modes which become critical
at the transition. If the marginal interaction between these modes was absent,
the structure factor at small $q$ would display a broad continuum
of states with a threshold $2|m_d|$\cite{snt}:
\be
S^{zz}(\omega,q,0) \sim \frac{q^2 m^2 _d}{s^3 \sqrt{s^2 - 4 m^2 _d}}.
\label{szz01}
\ee
As opposed to the case of the energy-density correlator (\ref{ed-ed-bos})
where the free-fermion approximation correctly captures its universal features,
here the interaction in the doublet sector changes the result (\ref{szz01})
dramatically. As already mentioned, apart from soliton/antisoliton states
the spectrum of the SGM (\ref{sg}) with $\frac{3}{4}\leq K\leq 1$ also
features the first breather state. The latter is odd under
charge conjugation ${\bf C}$, which inverts the sign of sine-Gordon
field $\Phi$ (see Eq.(\ref{parity1})). As follows from (\ref{br1}) the
current operator $I^3 = \sqrt{K/\pi} \p_x \Phi$ is also odd under
${\bf C}$ and consequently has a nonzero formfactor between the vacuum
and the first breather state. 
This gives rise to a coherent delta-function peak appearing
below the threshold of the two-soliton scattering continuum. 
Taking into account only the contributions
of the first breather (with mass $M_1$) and the two-particle scattering states,
close to the critical point we obtain
\bea
&&S^{zz}(\omega,q,0)
\propto 2\bigl|\frac{\pi\xi\lambda}{\sin(\pi\xi)}\bigr|^2
q^2 \delta(s^2 - M^2 _1)
\nonumber\\
&&+{4\pi}
\frac{q^2\sqrt{s^2 -4M^2 _d}}{s^3}
\frac{1}{\cosh(2\theta_0/\xi)+\cos(\pi/\xi)}\nn\\
&&\times
\exp\left(\int_0^\infty\!\frac{dt}{t}\frac{\sinh([1-\!\xi]t)[1-\cosh 2t
\cos(2\theta_0 t/\pi)]}{\sinh(2t)\ \cosh t\ \sinh(t\xi)}\right)\nn\\
\label{szz0ff}
\eea
where $\theta_0={\rm arccosh}(s/2M_d)$ and
\bea
\lambda&=&2\cos({\pi\xi}/{2})\sqrt{2\sin(\pi\xi/2)}\
\exp\left(-\int_0^{\pi\xi}
\frac{dt}{\sin t}\right)\ ,\nonumber\\
\xi&=&\frac{K}{2-K}\ ,\quad
M_1=2M_d\sin(\pi\xi/2)\ .
\label{notations}
\eea
This result is valid for any sign of $h-h_c$. We note
that $S^{zz}(\omega,q,0)$ vanishes as $q \to 0$ because the
$z$-component of the total magnetization, $S^z _1 + S^z _2 = \int \rd
x~I^3 (x)$, is conserved. 
In the limit $K\to 1$ the two-particle contribution to (\ref{szz0ff})
reduces to the form (\ref{szz01}) as it should, if we identify $M_d$
with $m_d$.

Exactly at the Gaussian criticality ($h=h_c$) the longitudinal structure factor
is given by the well-known ``chiral anomaly'' formula (see e.g. Ref.\cite{gnt}):
\be
S^{zz}(\omega,q,0) = K (\alpha/v) (q v)^2 \delta(\omega^2 - q^2 v^2).
\ee

\subsubsection{Transverse structure factor}
\label{sssec:TSF}

Now we turn to dynamical correlations of the transverse total current,
$I^1 = - \ri \xi^2 _{\nu} \xi^3 _{\nu}$.
As before
we will first adopt the free-fermion approximation
and then discuss how the results are affected by the interaction
between the doublet modes.

A simple calculation leads to the following expression for $S^{xx}(\omega,q,0)$
(we assume that $\omega>0$)
\bea
&&S^{xx}(\omega,q,0) \nn\\
&&= \frac{\alpha}{2v}
\frac{m^2 _+ q^2 (s^2 - m^2 _-) + m^2 _- (s^2 + q^2) (s^2 - m^2 _+)}
{s^4 \sqrt{(s^2 - m^2 _+)(s^2 - m^2 _-)}},
\label{final-formal}\\
&& m_{\pm} = m_3 \pm m_d, ~~s^2 \geq {\rm max}~ \{ m^2 _+, m^2 _- \}. \nn
\eea

\begin{figure}[ht]
\begin{center}
\epsfxsize=0.45\textwidth
\epsfbox{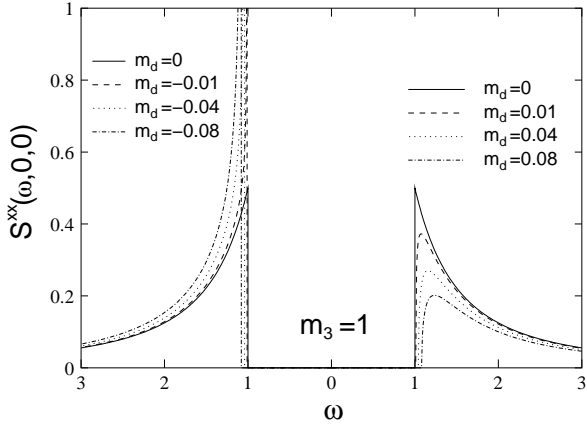}
\end{center}
\caption{Transverse structure factor at $q=q_{\perp}=0$.} 
\label{fig:fig3}
\end{figure}

The case $q=0$ is shown in Fig.\ref{fig:fig3}. Apart from the
$1/\omega^2$ decay at $\omega > m_3$, the behavior of $S^{xx}(\omega,q, 0)$
close to the threshold is similar to that of $S^{xx}(\omega,\pi+q, 0)$
(c.f. Fig.\ref{fig:fig2}).

\begin{figure}[ht]
\begin{center}
\epsfxsize=0.45\textwidth
\epsfbox{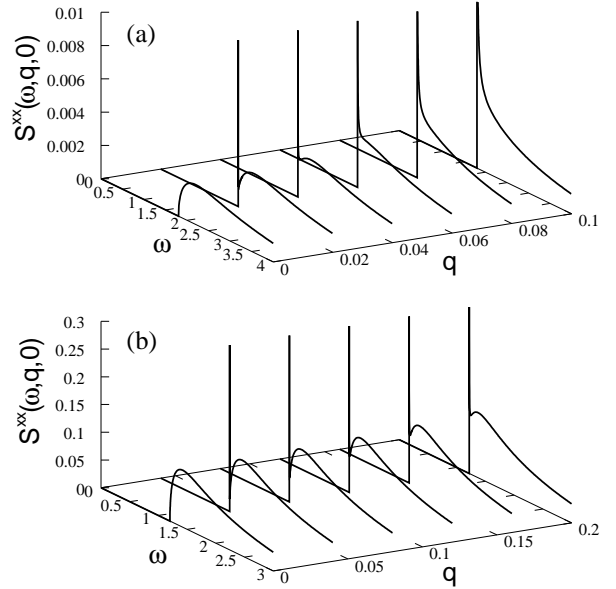}
\end{center}
\caption{Transverse structure factor at small $q$.\newline
a) $m_d = 0.8 m_3$; b) $m_d = 0.2 m_3$.} 
\label{fig:fig4}
\end{figure}

Although at $q=0$ the transverse structure factor is finite for
$h<h_c$ ($m_d \geq 0$), at arbitrarily small $q$ it is divergent
at the threshold (see Fig.\ref{fig:fig4}) 
\be
S^{xx}(\omega,q,0) \sim \frac{q^2 (m^2 _+ - m^2 _-)^{1/2}}{m^2 _+ (m^2 _+ + q^2)^{1/2}}
\frac{1}{\sqrt{\delta}},
\label{q-neq-0:sing}
\ee
where $\delta = \omega - \sqrt{m^2 _+ + q^2} > 0$.

For $h<h_c$ and at $q \neq 0$ the curves shown in Fig.\ref{fig:fig4} display an
interesting feature.  The $1/\sqrt{\delta}$ drop of the structure
factor slightly above the threshold is followed by an upturn which 
is a property of $S^{xx}(\omega,0,0)$. The maximum  occurs at
\[
\delta \sim \frac{(m^2 _+ - m^2 _-)q^2}{m_+ m^2 _-}.
\]

At $h>h_c$ ($m_d < 0$) the singularity at the threshold is of the form:
\be
S^{xx}(\omega, q,0) \sim \frac{(m^2 _- - m^2 _+)^{1/2} (m^2 _- + q^2)}{m^2 _-}
\frac{1}{\sqrt{\delta}}.
\ee
Now it survives the limit $q\to 0$ and disappears only at the critical point.

The free-fermion
approximation is reliable as long as $h \neq h_c$. 
The derivation of the structure of $S^{xx}
(\omega,q,0)$ at criticality is given in Appendix \ref{app-transverse}. 
Using (\ref{X6})-(\ref{relation-gamma}) we find that just above
the threshold $\omega = \sqrt{q^2 + m^2 _3}$ the structure factor
$S^{xx} (\omega, q,0)|_{\rm crit}$ only differs from the expression
(\ref{LL-threshold1}) by an extra factor $(1 + 2q^2 v^2 /m^2 _3)$
\bea
&&S^{xx} (\omega, q,0)\biggr|_{\rm crit}\nn\\
&&= {\cal C}(\theta) 
\left(1 + \frac{2q^2 v^2}{m^2 _3}  \right)
\left( \frac{s^2 - m^2 _3}{m^2 _3} \right)^{2\theta}.
\label{LL-threshold}
\eea
The discontinuity of $S^{xx} (\omega, q,0)$ at the threshold,
obtained in the free-fermion approximation,  is rounded by
the interaction in the doublet sector. However, due to the smallness
of $\theta$, the transverse structure factor rapidly increases beyond
the threshold and reaches values of the order of 1 exponentially close
to the threshold, $\omega - \sqrt{q^2 v^2 + m^2 _3} \sim \alpha^{-1}
e^{-1/2\theta}$. 

\subsection{Summary of the structure of the magnetic excitation
spectrum at weak coupling}
At this point it would appear to be useful to briefly summarize our
results concerning the behavior of the dynamical structure factor in
the weak-coupling regime derived above.
We do this in a simple pictorial way, having in mind to
illustrate the qualitative changes in the spin excitation
spectrum associated with the transition induced by the staggered
magnetic field.

\begin{figure}[ht]
\begin{center}
\epsfxsize=0.45\textwidth
\epsfbox{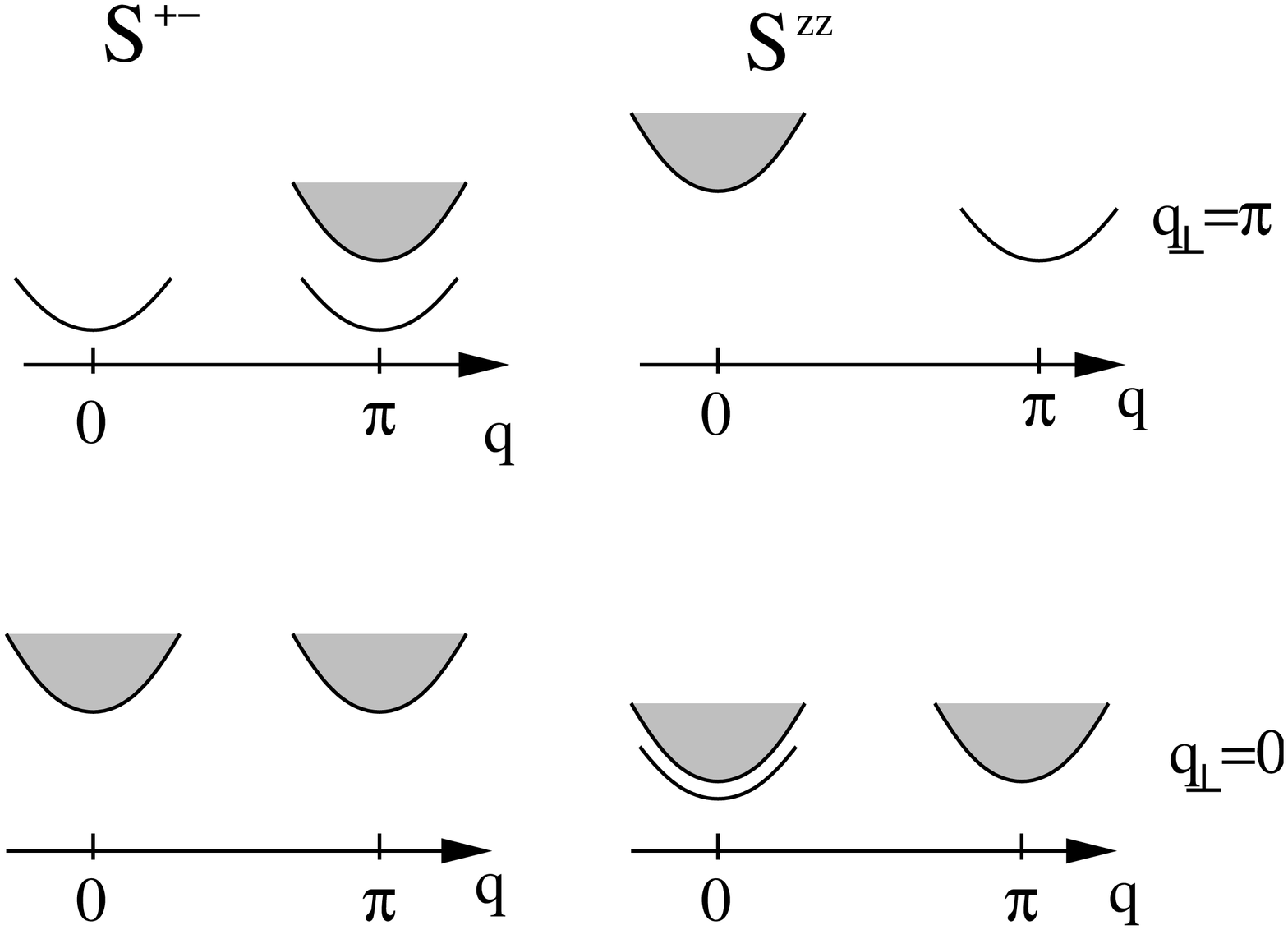}
\end{center}
\caption{\label{fig:weaklow} Structure of low-energy magnetic
excitations in the weak coupling limit and $h<h_c$.}
\label{fig:fig5}
\end{figure}

\begin{figure}[ht]
\begin{center}
\epsfxsize=0.45\textwidth
\epsfbox{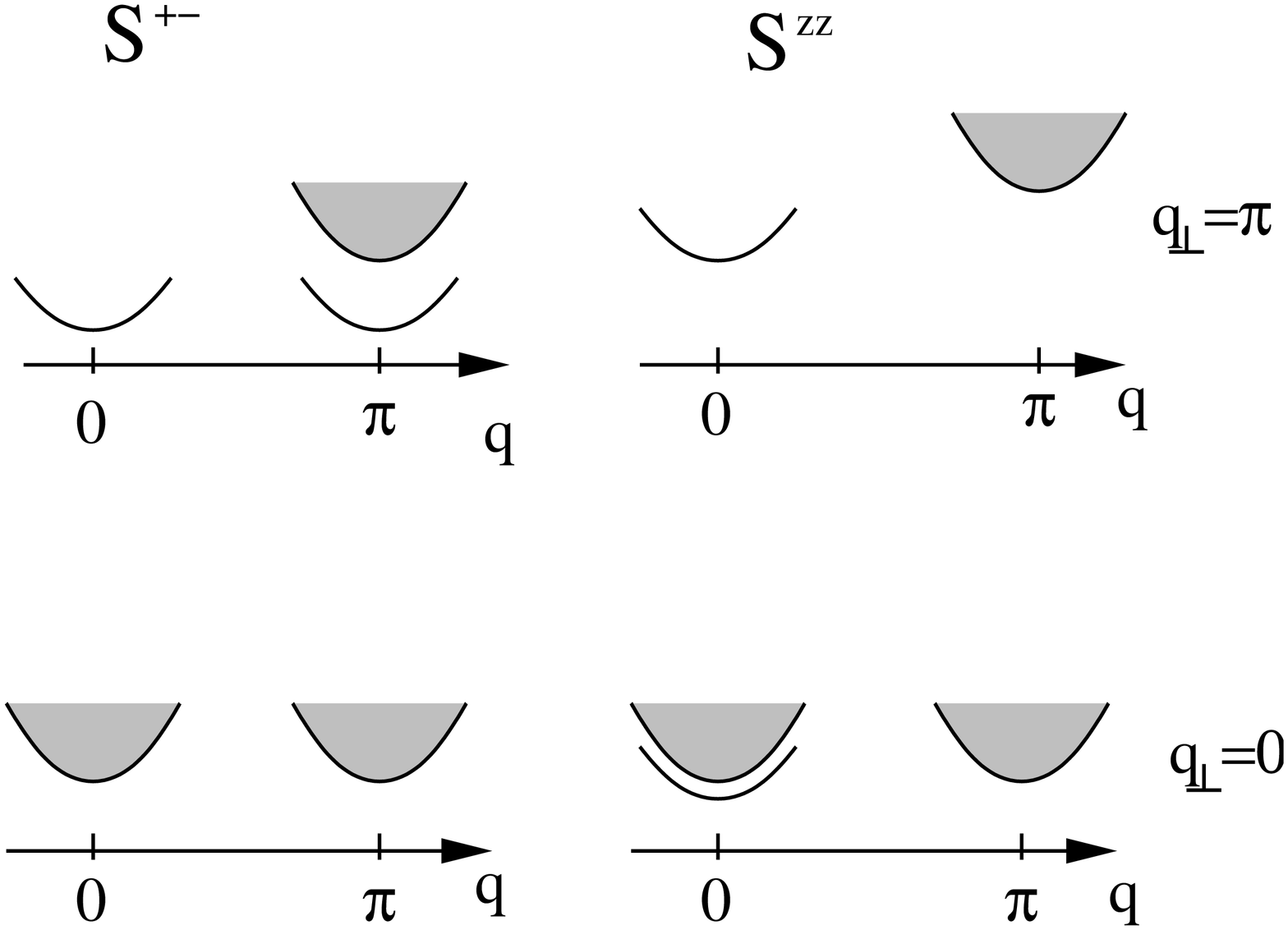}
\end{center}
\caption{\label{fig:weakhigh} Structure of low-energy magnetic
excitations in the weak coupling limit and $h>h_c$.}
\label{fig:fig6}
\end{figure}

In Figs \ref{fig:weaklow} and \ref{fig:weakhigh} curves denote
coherent single-particle excitations whereas shaded areas correspond
to incoherent multiparticle scattering continua.
The main changes as we tune the staggered field through the transition
concern $S^{zz}(\omega,q,\pi)$. The coherent mode visible in
$S^{zz}(\omega,q\approx\pi,\pi)$ below the transition is relaced by an
incoherent scattering continuum for $h>h_c$. In the vicinity of $q=0$
exactly the opposite happens: the incoherent scattering continuum
present below $h_c$ splits off a coherent mode for $h>h_c$.
The transverse magnon mode visible in $S^{+-}(\omega,q,\pi)$ is
well-defined both below and above the transition.

At $q_\perp=0$ the general structure of the excitation spectrum is
similar for $h<h_c$ and $h>h_c$. However, the precise from of the
structure factor changes significantly, which of course has important
consequences for the neutron scattering cross section.

\subsection{Structure factor at $|m_s| \ll h \ll J$}

This is the limit of two decoupled Heisenberg chains in a staggered
magnetic field, described a pair of the $\beta^2 = 2\pi$ sine-Gordon
Hamiltonians (\ref{sg-2pi}). The problem basically reduces to the
situation studied in \cite{oshikawa1,essler1,essler2,oshikawa2}.
The elementary excitations are
solitons ($s$) and antisolitons ($\bar{s}$) 
with a mass gap $M$ (which scales as $M \sim h^{2/3}$)
and two soliton-antisoliton ``breather'' bound states $B_{1,2}$
with gaps $M$ and $\sqrt{3}M$ respectively. The dynamical structure
factor has already been partially calculated by means of the
formfactor bootstrap approach in \cite{essler1}. We briefly review some
impotant ingredients of this method in section
\ref{sec:strongcoupling}. The leading contributions to the various
components of the structure factor are found to be
\bea
\label{t1}
&&S^{+-}(\omega,\pi+q,q_\perp)=A_\perp\ 
\delta(s^2 -M^2)+\ldots\ ,\\
\label{l1}
&&S^{zz}(\omega,\pi+q,q_\perp)=A_\parallel\ 
\delta(s^2 -3M^2)+\ldots\ ,\\
\label{t2}
&&S^{+-}(\omega,q,q_\perp)=B_\perp\
\frac{\omega^2}{M^2}\ 
\delta(s^2-M^2)+\ldots\ ,\\
\label{l2}
&&S^{zz}(\omega,q,q_\perp)=B_\parallel\ 
\frac{q^2}{M^2}\
\delta(s^2-M^2)+\ldots\ .
\eea
Here $A_{\perp,\parallel}$ and $B_{\perp,\parallel}$ are constants
that can be calculated, $q_\perp=0,\pi$ and the dots denote higher-energy
contributions of multiparticle intermediate states. 
Both at $q\sim 0$ and $q\sim \pi$
the transverse structure factor shows a coherent
delta function peak due to one-soliton intermediate states. 
This suggests that there may be a coherent transverse
soliton mode throughout the whole Brillouin zone.
The longitudinal structure factor exhibits a coherent delta-function
mode corresponding to the second breather $B_2$ with gap $\sqrt{3}M$
in the vicinity of $q=\pi$. For small momentum transfer along the
chain direction the longitudinal structure factor exhibits a coherent
delta-function peak which is a contribution of the first breather $B_1$ with
gap $M$. However, the spectral weight is very small ($\propto q^2$).

Let us now compare these results to the ones found above for the weak
coupling limit $J\gg J_\perp\gg h>h_c$. 
\medskip

\underline{$q_\perp=\pi$:}
The transverse stucture factor for $q\approx\pi$ is
similar to (\ref{t1}). The longitudinal structure factor at weak
coupling is incoherent whereas for $h\gg J_\perp$ it is coherent. This
is easily understood: increasing the staggered field eventually splits
off a bound state from the incoherent continuum. 
Finally, the structure factors for $q\approx 0$ have the same
general form as (\ref{t2}) and (\ref{l2}) respectively.
\medskip

\underline{$q_\perp=0$:}
Here the situation at large staggered fields, $h\gg J_\perp$, is
different from that found at $J_\perp\gg h>h_c$. The
transverse structure factor around $q=\pi$ and $q=0$ exhibits coherent
modes for $h\gg J_\perp$, Eqs. (\ref{t1}),(\ref{t2}), whereas it displays
incoherent scattering continua for $J_\perp\gg h>h_c$. However these
continua are singular above the thresholds, and one can thus easily
imagine that they are prone to split off coherent modes once $h$ becomes
sufficiently large. An analogous situation is encountered for
$S^{zz}(\omega,q,0)$ at small $q$. Finally, 
a coherent delta-function peak is found in 
$S^{zz}(\omega,\pi+q,0)$ both at
$h\gg J_\perp$ and $J_\perp\gg h>h_c$.


\section{Induced staggered magnetization}


In this section, we estimate the magnetic field dependence of the induced
staggered magnetization, ${\cal M}$, in the Haldane spin-liquid phase
of the spin ladder.
Since at $h=0$ correlations of the total staggered magnetization
$(-1)^n [S^z _{1n} + S^z _{2n}]$ are short-ranged, the dependence 
${\cal M} (h)$
will be linear in the limit
$h \to 0$: 
\bea
&& {\cal M}(h) = \chi_s (0) h, ~~~(h \to 0)\nn\\
&& \chi_{s} (0) = \frac{1}{v} \int \rd^2 {\bf r} \la
n^+ _z ({\bf r} )n^+ _z ({\bf 0} )\ra_0 
\sim \frac{J}{vJ_{\perp}}. \label{chi-0}
\eea
On the other hand, when $h$ is large, ($h \gg J_{\perp}$), the 
transverse coupling between the two S=1/2
Heisenberg chains of the
ladder can be neglected.  Each chain is then described at low energies
in terms of the $\beta^2 = 2\pi$ SGM (\ref{sg-2pi}). This leads to the result
\be
 {\cal M}(h)\propto \left( \frac{h \alpha}{v} \right)^{1/3},~~~~
\chi_{\rm stag} (h) \propto h^{-2/3}.
\ee

To estimate the singular part of 
of ${\cal M} (h)$ at $|h-h_c| \to 0$,
it is sufficient to know the $h$-dependence of the ground state energy density,
${\cal E} (h)$,
of the low-energy effective Hamiltonian. 
Both in the weak-coupling 
and strong-coupling limits, the latter has the structure of the SGM
(\ref{sg}) with the amplitude of the cosine term proportional to
$|h - h_c|$. Using standard scaling arguments, we find that
\[
{\cal E} (h) - {\cal E} (h_c) \sim - |h - h_c|^{\frac{2}{2-K}},
\]
where $K$ is given by Eq. (\ref{K-via-Delta}).
Higher-energy degrees of freedom which asymptotically decouple from the
critical ones provide a finite contribution
to ${\cal M}(h)$ at $h=h_c$. So
\bea
&&{\cal M}(h)=  {\cal M}(h_c) - 
\frac{\p {\cal E}(h)}{\p h}\nn\\
&&= m_{\rm stag} (h_c) +  {\rm const~}|h - h_c|^{\frac{K}{2-K}} {\rm sign~}(h-h_c).
\label{stag.aver.n}
\eea
The staggered susceptibility
\be
\chi_{s} (h) \sim |h-h_c|^{- \nu}, ~~~\nu = \frac{2(1-K)}{2-K} > 0,
\label{stag-chi}
\ee
is divergent at the transition. 
The nonuniversal exponent $\nu$ varies from $g_{\parallel}^c /\pi v$ at
$J_{\perp} \ll J$ to values close to 2/5 at $J_{\perp} \gg J$.


\section{Strong-coupling limit}
\label{sec:strongcoupling}. 

In this section, we consider the model (\ref{model}) in the
strong-coupling limit, 
$
J_{\perp}, h \gg J.
$
It will be assumed that the number of lattice sites is even, $N = 2M$,
and periodic boundary conditions are imposed.

In the $0^{\rm th}$-order approximation ($J = 0$), the Hamiltonian
describes a collection of even and odd rungs,
\be
H_0 = \sum_{m=1}^{M} H^0 _{2m} +  \sum_{m=0}^{M-1} H^0 _{2m+1},
\label{H0}
\ee
where
\be
H^0 _n = J_{\perp}{\bf S}_{1,n} \cdot {\bf S}_{2,n}
- h (-1)^n [S^z _{1,n} + S^z _{2,n}]. \label{rung-ham}
\ee
The spectrum of the even-rung Hamiltonian $H^0 _{2m}$ consists of 4 levels,
\bea
E_{\pm} &=& \frac{J_{\perp}}{4} \mp h,\quad
E_0 = \frac{J_{\perp}}{4},\quad E_s = - \frac{3J_{\perp}}{4}\ ,
\label{rung-spectrum}
\eea
corresponding to the triplet and singlet states
\bea
&&|+\ra = |\up,\up\ra,\
|0\ra = \frac{1}{\sqrt{2}} 
\biggl[|\up,\down\ra + |\down, \up\ra \biggr],\
|-\ra = |\down, \down\ra, \nn\\
&&|s\ra = \frac{1}{\sqrt{2}} 
\left(|\up, \down\ra - |\down, \up\ra \right). \nn
\eea
At $h < J_{\perp}$, the singlet state $|s\ra$ is the lowest-energy state.
At $h = J_{\perp}$ the $|+\ra$ level crosses with the singlet $|s\ra$
and at $h > J_{\perp}$ becomes the lowest energy state. Focussing on
the vicinity of the level crossing point ($|h - J_{\perp}| \ll
J_{\perp}$), we will retain two low-energy states on each even rung
\be
|\Up\ra_{2m} = |+ \ra_{2m}, ~~~|\Down\ra_{2m} = |s \ra_{2m}.
\label{newlabel1}
\ee
Projecting the even-rung Hamiltonian onto the subspace of these two states,
i.e. imposing the constraint\\
$
|\Up \ra \la\Up | + |\Down \ra \la\Down | = 1,
$
we obtain:
\bea
&&P H^0 _{2m} P = - \frac{1}{2} \left(\frac{J_{\perp}}{2} + h
\right)\nn\\
&&
- \frac{1}{2} \left(h - J_{\perp} \right)
\biggl\lbrace |\Up \ra_{2m} \la\Up |_{2m} -  |\Down \ra_{2m} \la\Down |_{2m}\biggr\rbrace.
\label{even-reduced}
\eea
The spectrum of the odd-rung Hamiltonian, $H^0 _{2m+1}$, is obtained from
(\ref{rung-spectrum})  by inverting the sign of $h$, which amounts
to interchanging the states $|+\ra$ and $|-\ra$. Thus the lowest-energy
states on odd rungs are
\be
|\Up\ra_{2m+1} = |s \ra_{2m+1}, ~~~|\Down\ra_{2m+1} = |- \ra_{2m+1},
\label{newlabel2}
\ee
and the projected odd-rung Hamiltonian reads
\bea
&&P H^0 _{2m+1} P = - \frac{1}{2} \left(\frac{J_{\perp}}{2} + h  \right)\nn\\
&&+ \frac{h - J_{\perp}}{2} \!
\biggl\lbrace |\Up \ra_{2m+1} \la\Up |_{2m+1} -  |\Down \ra_{2m+1}
\la\Down |_{2m+1}\biggr\rbrace.\! 
\label{odd-reduced}
\eea
Introducing effective spin-1/2 operators ${\bf T}_n$ associated
with the $n$-th rung,
\bea
&&T^z _{n} = \frac{1}{2} 
\biggl( |\Up \ra_{n} \la\Up |_{n} -  |\Down \ra_{n} \la\Down |_{n} \biggr),
\nn\\
&&T^+ _n = |\Up \ra_{n} \la\Down |_{n}, ~~~
T^- _n = |\Down \ra_{n} \la\Up |_{n},
\label{T-ops}
\eea
we see that, in the zeroth order in $J$, the low-energy Hamiltonian
describes a collection of noninteracting spins 1/2 on a 1d lattice in an 
effective staggered magnetic field $h - J_{\perp}$:
\be
PH_0 P = {\rm const~} - (h - J_{\perp}) \sum_n (-1)^n T^z _n .
\label{proj-H0}
\ee

The exchange interaction between the spins ${\bf T}_n$ is mediated by
the longitudinal ($J$) part of the Hamiltonian (\ref{model}).
In the two-dimensional low-energy subspaces of the even and odd rungs,
the original spin operators ${\bf S}_{j,n}$$(j=1,2)$ 
reduce to
\be
S^z _{j,n} \to \frac{1}{2} \left[ T^z _n + \frac{1}{2} (-1)^n \right],~~ 
S^{\pm}_{j,n} \to (-1)^{j+n} \frac{1}{\sqrt{2}} T^{\pm}_n .\label{S-vs-T}
\ee
As a result, in the first order in $J$, the interaction between
neighbouring rungs e.g. $2m$ and $2m+1$ is given by
\[
-J \left( T^x _{2m} T^x _{2m+1} +  T^y _{2m} T^y _{2m+1}
- \frac{1}{2} T^z _{2m} T^z _{2m+1} \right).
\]
It is convenient to make a unitary transformation
\be
T^{\pm} _n \to (-1)^n T^{\pm} _n, ~~ T^z _n \to T^z _n,
\ee
under which formul\ae\ (\ref{S-vs-T}) transform to
\bea
&&S^z _{1n} + S^z _{2n} \to T^z _n + \frac{1}{2} (-1)^n,  ~~S^z _{1n} - S^z _{2n} \to 0 \nn\\
&&S^+ _{1n} + S^+ _{2n} \to 0, ~~S^+ _{1n} - S^+ _{2n} \to - \sqrt{2} T^+ _n.
\label{transmut-sc}
\eea
The low-energy effective Hamiltonian takes the form of an anisotropic
spin-1/2 Heisenberg chain in a staggered magnetic field
\bea
H_{\rm eff} &=& {\rm const} + J \sum_n
\left( T^x _n T^x _{n+1} + T^y _n T^y _{n+1}
+ \Delta  T^z _n T^z _{n+1}\right) \nn\\
&&- h^* \sum_n (-1)^n T^z _n , \label{eff-final}
\eea
where
\be
\Delta = \frac{1}{2}, ~~~h^* = h - J_{\perp} + \frac{J}{2}. \label{param-eff}
\ee
Since $|\Delta| < 1$, the model (\ref{eff-final}) has a U(1)
(Gaussian) critical line $ h^* = 0$, i.e.
\be
h = J_{\perp} - \frac{J}{2} + O(J^2/J_{\perp}). \label{sc-crit.line}
\ee

To be sure of this result, we must verify that higher-order terms
($\propto J^2/J_{\perp}$),
originating from virtual transitions between the low-energy and
high-energy states, do not introduce relevant perturbations
at the U(1) criticality but only
renormalize the parameters of $H_{\rm eff}$ in (\ref{eff-final}).
Since the original Hamiltonian (\ref{model}) is site-parity symmetric,
a bond-alternating term cannot appear. Small corrections of the
order $J^2/J_{\perp}$ cannot make $\Delta$ greater than 1 and drive
the XXZ chain to the massive Neel phase. Next-nearest neighbour
exchange interactions with small coupling constants are also known to
be marginally irrelevant at low energies.
Finally, with the definitions (\ref{newlabel1}), (\ref{newlabel2})
of the spin-up and spin-down low-energy states,  the effective model
should be invariant under spin rotations around the staggered magnetic
field, which is a property of the original model.
This rules out a breakdown of the XY symmetry of the effective
Hamiltonian, which would generate a finite gap.
Thus, higher-order terms will slightly (in the order $J^2/J_{\perp}$) modify
the parameters of the effective Hamiltonian (\ref{eff-final}) and,
in particular, the critical line (\ref{sc-crit.line}), without affecting
the criticality itself. Small corrections to the zero-order 
parameter $\Delta = 1/2$ will keep the U(1) criticality far enough
from the SU(2) critical point $\Delta = 1$.


The picture of the transition emerging in the strong-coupling case,
$J_{\perp}, h \gg J$, is in full agreement with the one obtained 
in the weak-coupling limit, $J_{\perp}, h \ll J$ (see section II).
In both cases, the criticality is identified as that of an effective
spin--1/2 XXZ chain. 
We therefore expect that the existence of the U(1) criticality in the
antiferromagnetic S=1/2 two-leg ladder in a staggered magnetic field
is a {\sl universal} property of this system. 
The vicinity of the critical field is described by 
the SGM (\ref{sg}) with
\be
K = \frac{1}{2 (1 - \frac{1}{\pi} \arccos \Delta)}.
\label{K-via-Delta}
\ee
However, the equation for the critical line, 
as well as the value of $\Delta$ in the effective S=1/2 XXZ chain,
are sensitive to the strength of the
interchain coupling $J_{\perp}$. Being small at $J_{\perp} \ll J$, $\Delta$
increases with $J_{\perp}$ and tends to 1/2 
($K \to 3/4$) in the strong-coupling limit.
The scaling law for the critical line, $h_c \sim J^{3/2}_{\perp}$, valid 
for a weakly coupled ladder, can be easily obtained by comparing the mass gaps in the 
limiting cases $h /J_{\perp} \to 0$ and $J_{\perp}/h \to 0$. In the former case
the mass gap is linear in $J_{\perp}$
(up to logarithmic corrections). In the second case we have two decoupled
S=1/2 Heisenberg chains in a weak staggered magnetic field. Since the staggered
magnetization has scaling dimension 1/2, the mass gap scales as
$h ^{\frac{1}{2-d}} = h ^{2/3}$. The condition $J_{\perp} \sim  h^{2/3}$
brings us to (\ref{crit-curve1}). On increasing $J_{\perp}$, 
the power law $h_c \sim J^{3/2}_{\perp}$ gradually transforms to a linear dependence
(\ref{sc-crit.line}), valid at $J_{\perp} \gg J$
where the transition is governed by level crossing of the lowest-energy
on-rung spin states.

The strong-coupling approach can be applied to the generalized ladder model
(\ref{gen.model}) as well, provided that $|V| \ll J_{\perp}, h$.
The effective low-energy Hamiltonian is again of the form (\ref{eff-final}),
but its parameters are modified
\bea
&&J \to \bar{J} = J - \frac{V}{8}, ~~~ \Delta = 
\frac{1- \frac{3V}{8J}}{2(1- \frac{V}{8J})},\nn\\
&& h^* = h - J_{\perp} + \frac{J}{2} - \frac{V}{16}.
\label{par-gen}
\eea


\section{Physical properties in the strong-coupling limit}
\label{PPstrongcoupling}
In the weak-coupling case $J_{\perp} \ll J$ discussed above,
the exact relationship between the parameters of the low-energy model
of four massive Majorana fermions and those characterising the
original lattice spin ladder is unknown. For this reason, although the
weak-coupling approach correctly captures the universal parts of all
physical quantities in the vicinity of the critical point, 
the nonuniversal prefactors cannot be reliably estimated.
On the other hand, in the strong-coupling limit ($J_{\perp} \gg J$),
the mapping onto the effective XXZ spin-1/2 chain,
Eq.(\ref{eff-final}), is exact in the sense that all its parameters
can be found with any desired degree of accuracy in terms of the
expansion in powers of $J/J_{\perp}$.
Moreover, the projection the spin-ladder operators,
$S^{\alpha}_{j,n}$, onto those of the XXZ chain, $T^{\alpha}_n$, given
by Eqs.(\ref{transmut-sc}), does not contain nonuniversal parameters.
Therefore one can take advantage of the fact that the SGM
(\ref{sg}), describing the properties of the spin ladder in a 
staggered magnetic field close to its critical value, is integrable and
employ the formfactor bootstrap approach \cite{karo,smirnov,FF,lukPLA}
for a {\sl quantitative} analysis of the spectral properties of the system.

For simplicity we only consider the case $V=0$ but,
in view of Eqs.(\ref{par-gen}), the extension to a small nonzero $V$
is straightforward. 

In what follows we will use normalisations for the SGM
that are slightly different from those used in e.g. (\ref{sg-2pi}),
but are more convenient for the calculations we need to carry out.


\subsection{Scaling limit}

In the scaling limit, the XXZ spin chain with exchange constant $J$ and anisotropy
\be
\Delta=-\cos\pi\gamma^2
\ee
is described by the Gaussian model with Lagrangian
\bea
{\cal L}&=&\frac{1}{16\pi}(\partial_\mu\phi)^2 .
\eea
In the strong-coupling limit we have $\Delta=\frac{1}{2}$ and thus 
\be
\gamma^2=\frac{2}{3}.
\ee
The scaling limit is defined by
\be
J\to\infty\ , ~~ a_0\to 0\ ,~~
Ja_0 = 2\frac{1-\gamma^2}{\sin\pi\gamma^2}=\ {\rm fixed}. 
\label{fixing-v}
\ee
In these conventions $a_0$ is scaled in such a way that the spin
velocity is set to 1. It is easily restored in the final results by
dimensional analysis. Following Lukyanov \cite{lukPLA}, we will
normalize the field $\phi$ according to the short-distance OPE 
\bea
e^{i\gamma\phi(x)}\ e^{i\gamma\phi(y)}\
\longrightarrow |x-y|^{-4\gamma^2}\ ,\quad |x-y|\to 0.
\eea
Note that this implies that the lattice spacing must be taken into
account explicitly when relating lattice operators to field theory
ones. For example, for the staggered components of the spin operators
we have 
\bea
2\tt^\pm_n&\to&(-1)^na_0^\frac{\gamma^2}{2}\sqrt{\frac{F}{2}}\ \exp\left(\pm
i\frac{\gamma}{2}\theta\right)\ ,\nn\\
2\tt^z_n&\to&(-1)^na_0^\frac{1}{2\gamma^2}\sqrt{2A}\ \cos\left(\frac{1}{2\gamma}\phi\right)\ .
\eea
The {\sl nonuniversal} constants $A$ and $F$ are known exactly
\cite{coeffs}
\bea
&&F=\frac{1}{2(1-\gamma^2)^2}\left[\frac{\Gamma\left(\frac{\gamma^2}{2-2\gamma^2}\right)}
{2\sqrt{\pi}\Gamma\left(\frac{1}{2-2\gamma^2}\right)}\right]^{\gamma^2}\nn\\
&&\times\exp\left(-\int_0^\infty
\frac{dt}{t}\left[\frac{\sinh(\gamma^2t)}
{\sinh t\ \cosh(t[1-\gamma^2])}-\gamma^2e^{-2t}\right]\right),\nn\\
\eea
\bea
&&A=\frac{8}{\pi^2}\left[\frac{\Gamma\left(\frac{\gamma^2}{2-2\gamma^2}\right)}
{2\sqrt{\pi}\Gamma\left(\frac{1}{2-2\gamma^2}\right)}\right]^{\frac{1}{\gamma^2}}
\exp\Bigl(\int_0^\infty\frac{dt}{t}\Bigl[\nn\\
&&\frac{\sinh([2\gamma^2-1]t)}
{\sinh(\gamma^2 t)\ \cosh(t[1-\gamma^2])}
-(2-\frac{1}{\gamma^2})e^{-2t}
\Bigr]\Bigr).
\label{A}
\eea
For $\gamma^2=2/3$ we obtain $F\approx 0.5360$ and $A\approx
0.4285$. When the staggered field term is added, the Lagrangian 
density becomes 
\bea
{\cal L}&=&\frac{1}{16\pi}(\partial_\mu\phi)^2-2\mu\ \cos\b \phi\ ,
\label{sg-L}
\eea
where $\b=1/2\gamma=\sqrt{3/8}$ and for $h^*\leq 0$ 
\bea
2\mu&=&-\frac{h^*}{2}\sqrt{2A}a_0^{2\b^2-1}\ ,\nn\\
2\tt^\pm_n&=&(-1)^na_0^{1/8\b^2}\sqrt{\frac{F}{2}}\ \exp\left(\pm
i\frac{1}{4\b}\theta\right)\ ,\nn\\
2\tt^z_n&=&-(-1)^na_0^{2\b^2}\sqrt{2A}\ \cos\b\phi\ .
\label{mu}
\eea
For $h^*>0$ the signs of $\mu$ and of the expression for $\tt^z_n$
need to be inverted. 
The spectrum of the SGM (\ref{sg-L}) at $\b^2=3/8$
consists of soliton and antisoliton with gap $M_d$ and one
soliton-antisoliton bound state called ``breather'' with gap
\be
M_1=2M_d\ \sin(\pi\xi/2)\ ,~~~~\xi=\frac{\b^2}{1-\b^2}=\frac{3}{5}\ .
\label{1breather}
\ee
The soliton gap can be expressed in terms of the scale $\mu$ by
comparing the results of a Thermodynamic Bethe Ansatz calculation
with those of a perturbative calculation valid at high energies
\cite{zam1995}
\be
\mu=\frac{\Gamma(\b^2)}{\pi\Gamma(1-\b^2)}\left[M_d \frac{\sqrt{\pi}}{2}
\frac{\Gamma([1+\xi]/2)}{\Gamma(\xi/2)}\right]^{2-2\b^2}\ .
\label{lambdamu}
\ee
Combining (\ref{lambdamu}) with (\ref{mu}) we can express the
soliton gap $M_d$ in terms of the microscopic parameters of the lattice
model
\bea
&&\frac{M_d}{J}=\alpha\ \left(\frac{|h^*|}{J}\right)^{4/5}\ ,\nn\\
&&\alpha= \frac{3\sqrt{3}\Gamma(3/10)}{2\sqrt{\pi}\Gamma(8/10)}
\left(\frac{\pi\Gamma(5/8)\sqrt{2A}}{3\sqrt{3}\Gamma(3/8)}\right)^{4/5}\approx
1.58424 .
\eea
\subsection{Staggered Magnetization}
The staggered magnetization is given by
\begin{equation}
\langle (-1)^n\ S^z_{j,n}\rangle= \frac{1}{4}+\frac{1}{2}\langle
(-1)^n\ T^z_{n}\rangle\ .
\end{equation}
The point to note here is that, in order to be close to criticality, the
staggered field must be large, $h\approx J_\perp$, and consequently, the
staggered magnetization is large as well.
In the scaling limit, i.e. in the vicinity of the critical line
$h^*=0$, we can use the SGM to determine the deviation from $1/4$:
\begin{equation}
\langle
(-1)^n\ T^z_{n}\rangle={\rm sgn}(h^*)\ 
\frac{a_0^{2\b^2}}{2}\sqrt{2A}\ \langle
\exp(i\b\phi)\rangle\ .
\end{equation}
Expectation values of vertex operators have been determined in
Ref.\cite{lz97}, and in particular we have
\bea
\langle\exp(i\b\phi)\rangle&=&\frac{(1+\xi)\pi\Gamma(1-\b^2)}
{16\sin(\pi\xi)\ \Gamma(\b^2)}\nn\\
&\times&\left(\frac{\Gamma(\frac{1}{2}+\frac{\xi}{2})\Gamma(1-\frac{\xi}{2})}
{4\sqrt{\pi}}\right)^{2\b^2-2} M^{2\b^2}_d.
\label{expec}
\eea
Combining (\ref{expec}) with (\ref{fixing-v}) and (\ref{A}) we obtain
\begin{equation}
\langle(-1)^n\ T^z_{n}\rangle={\rm sgn}(h^*)\ 
0.2897\ \left(\frac{|h^*|}{J}\right)^\frac{3}{5}\ .
\end{equation}


\subsection{Dynamical Structure factor}


Our task is now to calculate the Fourier transform of the retarded
dynamical correlation functions in the SGM. This is done
by going to the spectral representation and then utilising the
integrability of the SGM to determine exactly the matrix elements of
the specific operator under consideration
between the ground state and various excited states. This method is
known as the form factor bootstrap approach \cite{smirnov}. 
Let us review some of its relevant steps.

In order to utilize the spectral representation, we need to specify a
basis of eigenstates of the Hamiltonian. Such a basis
is given by scattering states of breathers, solitons and antisolitons.
To distinguish these, we introduce labels $B,s,\bar{s}$. As usual,
for particles with relativistic dispersion it is 
convenient to introduce a rapidity variable $\theta$ to parametrize
energy and momentum 
\begin{eqnarray}
E_{s}(\theta)&=&M_d\cosh\theta\,~~P_{s}(\theta)=M_d\sinh\theta\
,\nonumber\\
E_{\bar{s}}(\theta)&=&M_d\cosh\theta\,~~P_{\bar{s}}(\theta)=M_d\sinh\theta\
,\nonumber\\
E_{B}(\theta)&=&M_1\cosh\theta\,~~P_{B}(\theta)=M_1\sinh\theta\ ,
\end{eqnarray}
where the breather gap $M_1$ is given above. 
A basis of the scattering states 
can be constructed by means of the Zamolodchikov-Faddeev (ZF)
algebra. The ZF algebra can be considered to be 
an extension of the algebra of creation and annihilation operators for
free fermion or bosons to the case of interacting particles with
factorizable scattering. 
This algebra is based on the knowledge of the exact spectrum and the
scattering matrix \cite{SM}. For the SGM
the ZF operators (and their hermitian
conjugates) satisfy the following algebra
\begin{eqnarray}
{Z}^{\epsilon_1}(\theta_1){Z}^{\epsilon_2}(\theta_2) &=&
S^{\epsilon_1,\epsilon_2}_{\epsilon_1',\epsilon_2'}(\theta_1 -
\theta_2){Z}^{\epsilon_2'}(\theta_2){Z}^{\epsilon_1'}(\theta_1)\ ,
\nonumber\\
{Z}_{\epsilon_1}^\dagger(\theta_1)Z_{\epsilon_2}^\dagger(\theta_2) &=&
Z_{\epsilon_2'}^\dagger(\theta_2){ Z}_{\epsilon_1'}^\dagger
(\theta_1)S_{\epsilon_1,\epsilon_2}^{\epsilon_1',\epsilon_2'}(\theta_1 -
\theta_2) , \nonumber\\
Z^{\epsilon_1}(\theta_1)Z_{\epsilon_2}^\dagger(\theta_2) &=&Z_{\epsilon_2'}
^\dagger(\theta_2)
S_{\epsilon_2,\epsilon_1'}^{\epsilon_2',\epsilon_1}
(\theta_2-\theta_1)Z^{\epsilon_1'}(\theta_1)\nonumber\\
&&+(2 \pi) \delta_{\epsilon_2}^{\epsilon_1} 
\delta (\theta_1-\theta_2).
\label{fz1}
\end{eqnarray}
Here $S^{\epsilon_1,\epsilon_2}_{\epsilon_1',\epsilon_2'}(\theta)$ are
the (factorizable) two-particle scattering matrices and
$\varepsilon_j=s,\bar{s},B$.

Using the ZF generators, a Fock space of states can be constructed as
follows. The vacuum is defined by
\begin{equation}
Z_{\varepsilon_i}(\theta) |0\rangle=0 \ .
\end{equation}
Multiparticle states are obtained by acting with strings of
creation operators $Z_\epsilon^\dagger(\theta)$ on the vacuum
\begin{equation}
|\theta_n\ldots\theta_1\rangle_{\epsilon_n\ldots\epsilon_1} = 
Z^\dagger_{\epsilon_n}(\theta_n)\ldots
Z^\dagger_{\epsilon_1}(\theta_1)|0\rangle . 
\label{states}
\end{equation} 
In term of this basis the resolution of the identity reads
\bea
&&\sum_{n=0}^\infty\sum_{\epsilon_i}\int_{-\infty}^{\infty}
\frac{d\theta_1\ldots d\theta_n}{(2\pi)^nn!}
|\theta_n\ldots\theta_1\rangle_{\epsilon_n\ldots\epsilon_1}
{}^{\epsilon_1\ldots\epsilon_n}\langle\theta_1\ldots\theta_n|\ \nn\\
&& =1 .
\label{identity}
\eea
Inserting (\ref{identity}) between operators in a 2-point correlation 
function
we obtain the following spectral representation 

\begin{eqnarray}
\label{corr}
&&\langle {\cal O}(x,t){\cal O}^\dagger(0,0)\rangle
=\sum_{n=0}^\infty\sum_{\epsilon_i}\int
\frac{d\theta_1\ldots d\theta_n}{(2\pi)^nn!}
\nonumber\\
&&\times
\exp\Bigl({\ri \sum_{j=1}^n p_jx-e_j t}\Bigr)
|\langle 0| {\cal O}(0,0)|\theta_n\ldots\theta_1
\rangle_{\epsilon_n\ldots\epsilon_1}|^2, 
\end{eqnarray}
where 
$p_j$ and $e_j$ are given by
\begin{equation}
p_j=M_{\epsilon_j}\sinh \theta_j, \; e_j=M_{\epsilon_j}\cosh \theta_j,
\label{epj}
\end{equation}
and 
\begin{equation}
\label{formf}
f^{\cal O}(\theta_1\ldots\theta_n)_{\epsilon_1\ldots\epsilon_n}\equiv
\langle 0| {\cal
O}(0,0)|\theta_n\ldots\theta_1\rangle_{\epsilon_n\ldots\epsilon_1} 
\end{equation}
are the form factors (FF). Our conventions in (\ref{epj}) are such that
$M_s=M_{\bar{s}}=M_d$ and $M_{B}=M_1$. 
After carrying out the double Fourier transform we obtain the
following expression for the dynamical structure factor
\begin{eqnarray}
\label{expansion1}
&&S^{\cal O}(\omega,q)= \frac{1}{2\pi} \int_{-\infty}^\infty dx\int_{-\infty}^\infty dt
\ e^{i\omega t-iqx} \langle
{\cal O}(x,t) {\cal O}^\dagger(0,0)\rangle \nonumber\\
&&= 2\pi\sum_{n=0}^\infty\sum_{\epsilon_i}\int
\frac{d\theta_1\ldots d\theta_n}{(2\pi)^nn!}
|f^{\cal O}(\theta_1\ldots\theta_n)_{\epsilon_1\ldots\epsilon_n}|^2 \nonumber\\
&&\qquad\times \
\delta(q-\sum_jM_{\epsilon_j}\sinh\theta_j)\
\delta(\omega-\sum_jM_{\epsilon_j}\cosh\theta_j)\ .
\nonumber\\
\label{strfa}
\end{eqnarray}
Let us now evaluate the leading contributions to (\ref{strfa}) in the
physically relevant cases. 

\subsection{Transverse structure factor at $q\approx \pi$}

To determine the transverse structure factor, we need 
FFs of the operator 
\begin{equation}
\exp(\frac{i}{4\b}\theta)\ .
\end{equation}
The lowest-lying states to which it couples are one-soliton states, and the 
first nonvanishing 
FF is a constant \cite{lz2001} 
\be
\langle
0|\exp\left(\frac{i}{4\b}\theta(0,0)\right)|\theta\rangle_-=\sqrt{Z_1(0)}\ .
\ee
In our case we have
\bea
&&Z_1(0)=\left[\frac{4}{\xi}\exp\left(\int_0^\infty
\frac{dt}{t}\frac{\sinh t\ \sinh(t[\xi-1])}{\sinh(t\xi)\ \cosh^2
t}\right)\right]^\frac{1}{4}\nn\\
&&\times
\left[\frac{\sqrt{\pi}\Gamma(\frac{3}{2}+\frac{\xi}{2})}
{\Gamma(\frac{\xi}{2})}\right]^{\gamma^2}M^{\gamma^2}_d\
\exp\Bigl(2\int_0^\infty
\frac{dt}{t}\Bigl[\frac{1}{4\sinh(t\xi)}\nn\\
&&\qquad +\frac{e^{(1+\xi)t}-1}
{4\sinh(\xi t)\ \cosh t\ \sinh([1+\xi]t)}
-\frac{\gamma^2\ e^{-2t}}{2}\Bigr]\Bigr)\nn\\
&&\approx 4.01\ M^{2/3}_d\ .
\eea
The corresponding contribution to the dynamical structure factor is
\bea
&&S^{+-}(\omega,\pi+q)=\frac{F}{4}Z_1(0)a_0^{2/3}\ \delta(s^2 -M^2 _d )\ ,\nn\\
&&\qquad\approx 0.614\ \left(\frac{|h^*|}{J}\right)^{8/15}\ 
\delta(s^2 - M^2 _d ) .\
\label{spm}
\eea
The dynamical susceptibility of the original ladder model can easily
be restored by means of the relations (\ref{transmut-sc}) between the
$\tt_n^\alpha$ variables and the original spin operators $S_{n,j}^\alpha$
\bea
S^{+-}_{\rm ladder}(\omega,\pi+q,\pi)&=&S^{+-}(\omega,\pi+q)\ ,\nn\\
S^{+-}_{\rm ladder}(\omega,\pi+q,0)&=&0.
\eea

The nice thing about this result is that we are able to calculate the
spectral weight in the delta-function {\sl exactly}.
This result is exact up to frequencies $\omega=3M_d$, where additional
contributions from $ss\bar{s}$ states arise. These can, in principle, be
calculated by first determining the corresponding 
FFs from
the relevant annihilation pole conditions (see e.g. \cite{karo} for a
similar calculation) and then by carrying out the remaining integrations
over the rapidity variables numerically.
The contribution of (\ref{spm}) to the total spectral weight at
$q=\pi$ is proportional to $\left(\frac{|h^*|}{J}\right)^{-\frac{4}{15}}$ and therefore
diverges as the field $h^*$ goes to zero. At first sight this
may look strange, but the same is true for the gapless XXZ chain,
where
\bea
S^{+-}_{\rm XXZ}(\omega,\pi)\propto\frac{1}{\omega^{2-\eta}}\ ,
\eea
where $\eta=1-\frac{1}{\pi}{\rm arccos}\Delta$.


\subsection{Longitudinal structure factor at $q\approx\pi$}

\begin{figure}[ht]
\begin{center}
\epsfxsize=0.45\textwidth
\epsfbox{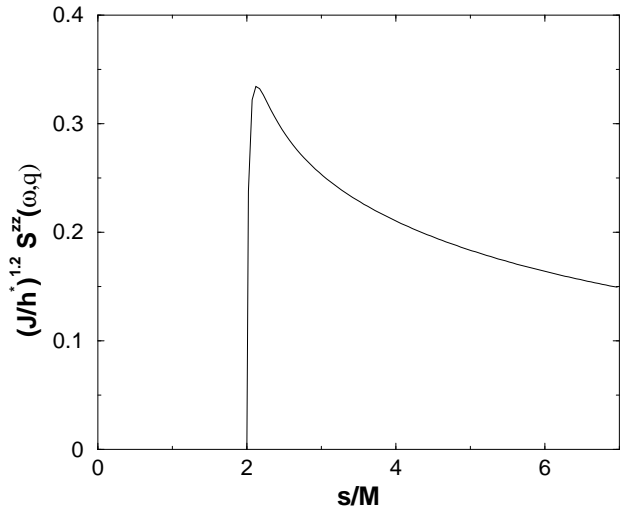}
\end{center}
\caption{\label{fig:szz} Longitudinal staggered structure factor
at $q_{\perp} = 0$.}
\end{figure}

In order to determine the longitudinal structure factor we need the
FFs of the operator $\cos\b\phi$. The first nonvanishing FF is between
the vacuum and two-particle $s\bar{s}$ states. The FF is given by
\cite{lukPLA} 
\bea
&&\langle 0|\cos\b\phi(0,0)|\theta_1,\theta_2\rangle_{-+}
={{\cal G}}_{\b}F(\theta_{12})\ ,\nn\\
&&F(\theta)=
\cot(\pi\xi/2)\frac{2i\cosh(\theta/2)}
{\xi\sinh([\theta+i\pi]/2\xi)}\sinh(\theta/2)\nn\\
&&\times
\exp \left ( \int_0^\infty\frac{dt}{t}
\frac{\sinh^2(t[1-i\theta/\pi])\sinh(t[\xi-1])}{\sinh 2t\
\sinh\xi t\ \cosh t}\right),\nn\\ 
&&{{\cal
G}}_{\b}=\left[\frac{M_d \sqrt{\pi}\Gamma(\frac{1}{2-2\b^2})}
{2\Gamma(\frac{\b^2}{2-2\b^2})}\right]^{2\b^2}\ \exp \left( 
\int_0^\infty\frac{dt}{t}f(t)\right),\nn\\
&&f(t)=\frac{\sinh^2(2\b^2 t)}{2\sinh(\b^2t)\
\sinh t\ \cosh([1-\b^2] t)}-2\b^2\ e^{-2t}\ ,\nn\\
\eea
where $\theta_{12}=\theta_1-\theta_2$. 
Carrying out the Fourier transformation, we obtain
\bea
&&S^{zz}(\omega,\pi+q)\simeq\frac{A}{2}a_0^{4\b^2}|{{\cal G}}_{\b}|^2
\frac{2}{\pi}\frac{1}{s\sqrt{s^2-4M^2 _d}}
|F(2\theta_0)|^2\nn\\
&&\qquad\approx1.13\ \left(\frac{|h^*|}{J}\right)^{6/5}\frac{2}{\pi}\frac{1}
{s\sqrt{s^2-4M^2 _d}}
|F(2\theta_0)|^2\ .
\eea
where $s^2=\omega^2-q^2$ and $\theta_0={\rm arccosh}(s/2M_d)$.
We plot the result in Figure \ref{fig:szz}.
Just above the threshold at $\omega=\sqrt{q^2+4M_d^2}$ the structure
factor increases in a universal square root fashion. This is easily
seen by considering the limit $\theta_0\ll 1$ of the function
$|F(2\theta_0)|^2$
\bea
|F(2\theta_0)|^2\propto\frac{s^2-4M_d^2}{M_d^2}\ ,
\eea

Using the relations (\ref{transmut-sc}) we finally obtain the
longitudinal structure factor of the original ladder model
\bea
S^{zz}_{\rm ladder}(\omega,\pi+q,0)&=&S^{zz}(\omega,\pi+q)\ ,\nn\\
S^{zz}_{\rm ladder}(\omega,\pi+q,\pi)&=&0.
\label{relationtoladder}
\eea
This result is exact for frequencies below $\omega=4M_d$, where
additional contributions due to intermediate states with 2 solitons
and 2 antisolitons arise. 
Comparing the transverse structure factor to the longitudinal one,
we see that, in the latter case, the contribution of the $s\bar{s}$
intermediate states to the total spectral weight is proportional to
$\left({|h^*|}/{J}\right)^\frac{2}{5}$ and thus becomes small at
small fields $h^*$. Thus at low energies the coherent soliton modes in
the transverse structure factor completely dominate the magnetic
response around the antiferromagnetic wave number $q=\pi$.

\subsection{Transverse structure factor at $q\approx 0$}
The smooth components of the transverse spin operators in the XXZ
chain are proportional to
\bea
&&{\cal J}^\pm=
\exp\left(\pm\frac{i}{4\bar{\beta}}\theta\mp i\bar{\beta}\phi\right)
+\exp\left(\pm\frac{i}{4\bar{\beta}}\theta\pm i\bar{\beta}\phi\right).
\label{ssmooth}
\eea
We note that the operator (\ref{ssmooth}) reduces to the sum of
the chiral SU(2) currents in the XXX case. Using the results of
\cite{lz2001} one can determine the leading contribution to the
transverse structure factor at small energies by means of the
formfactor bootstrap approach. The contribution of one soliton
intermediate states is given by
\bea
\langle {\cal J}^-(\tau,x)\ {\cal J}^+(0)\rangle&\propto&
K_0(M_d r)-\frac{x^2-v^2\tau^2}{x^2+v^2\tau^2}K_2(M_d r),\nn\\
\eea
where $r^2=x^2+v^2\tau^2$. Carrying out the Fourier transformation and
analytically continuing we arrive at the following result for the
dynamical structure factor
\bea
S^{-+}(\omega,q,\pi)={\rm const}\frac{\omega^2}{M^2 _d}\
\delta(s^2-M^2_d)+\ldots
\eea

\subsection{Longitudinal structure factor at $q\approx 0$}

The longitudinal structure factor around $q=0$ can be calculated
analogously. The relevant Fourier component of the spin operator
is given by
\be
T^z_n\simeq\frac{a_0\bar{\beta}}{2\pi}\ \partial_x\phi\ .
\ee
Here $\partial_x\phi$ is the topological charge density in the SGM and
its formfactors are known \cite{smirnov,karo,FF}. As we have mentioned before,
there exists a soliton-antisoliton bound state which gives rise to a
{\sl coherent} delta-function contribution to the longitudinal
structure factor around $q=0$. The contribution with the next highest
threshold in energy is due to a soliton-antisoliton scattering
continuum. Taking only these two contributions into account we obtain

\bea
&&S^{zz}(\omega,q)\simeq\frac{0.0617}{J^2}
\frac{q^2}{\sqrt{q^2+M_1^2}}\ 
\delta\left(\omega-\sqrt{q^2+M_1^2}\right)\nonumber\\
&&+\frac{16}{27\pi J^2}
\frac{q^2\sqrt{\omega^2-q^2-4M^2 _d}}{(\omega^2-q^2)^\frac{3}{2}}
\frac{1}{\cosh(2\theta_0/\xi)+\cos(\pi/\xi)}\nn\\
&&\times
\exp\left(\int_0^\infty\!\frac{dt}{t}\frac{\sinh([1-\!\xi]t)[1-\cosh 2t
\cos(2\theta_0 t/\pi)]}{\sinh(2t)\ \cosh t\ \sinh(t\xi)}\right),\nn\\
\label{szzq=0}
\eea
where we again have defined $\theta_0={\rm arccosh}(s/2M_d)$.
The relation of $S^{zz}$ to the structure factor of the original
ladder model is given by (\ref{relationtoladder}) with $\pi+q$ replaced 
by $q$. 
As expected the structure factor vanishes like $q^2$ as $q\to 0$. This
behaviour is completely fixed by the Lorentz invariance of the
low-energy effective theory and the fact that the topological charge
density is part of a Lorentz vector.
The result (\ref{szzq=0}) holds in the small momentum region and it
should be possible to resolve the structure of excitations in terms of the
breather bound state and the $s\bar{s}$ scattering continuum by
carrying out Neutron scattering experiments  at small momentum transfer.

\begin{figure}[ht]
\begin{center}
\epsfxsize=0.45\textwidth
\epsfbox{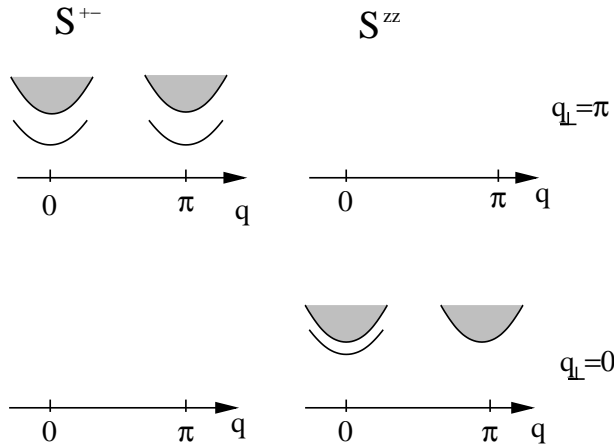}
\end{center}
\caption{\label{fig:strong} Structure of low-energy magnetic
excitations in the strong coupling limit. This picture is valid
both at $h<h_c$ and $h>h_c$.}
\end{figure}

Our results for the dynamical structure factor factor in the strong
coupling limit imply the structure of low-lying excited states shown
in Fig.\ref{fig:strong}.

\section{Spontaneously dimerized ladder}

Now we turn to the generalized spin-ladder model (\ref{gen.model}) in which
the four-spin interaction $V$ is superimposed on the 
antiferromagnetic exchange ($J_{\perp}>0$) across the rungs.
If $V$ is positive and large enough, the ladder occurs in a non-Haldane,
spontaneously dimerized phase with a fully incoherent spectrum
exhausted by pairs of massive dimerization kinks \cite{nt,km}.

The tendency towards suppression of the spin-liquid phase upon
increasing $V$ is already seen in the strong-coupling limit;
see formulas (\ref{par-gen}). Within the weak-coupling scheme ($J_{\perp}, V \ll J$),
the transition to the spontaneously dimerized phase is associated with the sign reversal
of the triplet mass. From Eqs. (\ref{n+}) it then follows that, when all  
four Ising copies are ordered (the case $m_{s}, m_t < 0$),
the ground state is dimerized and doubly degenerate, with 
$\la \eps^- \ra = \pm \eps_0$ being the order parameter,
whereas the spin excitation spectrum represents a broad continuum
with thresholds at $2|m_t|$ and $|m_t| + |m_s|$.

At the transition from the spin-liquid to the spontaneously dimerized phase, the Majorana
triplet $\vxi$ becomes massless ($m_t = 0$), and the 
system becomes critical. The criticality 
belongs to the universality class of the level $k=2$ SU(2) WZNW model with
central charge $C=3/2$ (see Ref.\cite{nt}). When a staggered magnetic field is applied,
then, in the parameter space ($m_s, m_t, h$), the 
semi-infinite critical line $m_t = 0, m_s < 0$
splits in the direction of the field $h$ into two critical surfaces:
one corresponding to the U(1) criticality with central charge $C=1$, already
considered in previous sections, and the other representing the Ising criticality
with $C=1/2$ (see Fig.\ref{fig:fig1}). The latter will be discussed in this section.

The existence of an Ising QCP in the generalized ladder model
can be understood using an argument similar to the
one in section II. In the spontaneously dimerized phase all Ising copies are ordered
($m_t, m_s < 0$). Then, in the leading order, the interaction term in (\ref{total-ham-dens})
can be replaced by $\tilde{h} \mu_3 \mu_4$, where $\tilde{h}\sim \hat{h}\la \s_1 \s_2 \ra$.
If the  masses $m_3 = m_t$ and $m_4 = m_s$ were equal, the resulting model would
be equivalent to the double-frequency SGM in which an Ising QCP has
already been described in much detail\cite{dm,fgn}. The
existence of this transition can be easily visualized in the strong-coupling limit 
(large $\tilde{h}$).
In this limit, the ``relative'' Ising degree of freedom, $\nu = \mu_3 \mu_4$, becomes effectively
frozen out, while the ``total'' degrees of freedom, described, e.g., by $\mu_3$ can be tuned
to criticality. This argument is still valid if the two Ising systems have different mass gaps,
provided that they are in the same (in this, case, ordered) phase.

Starting from the spontaneously dimerized phase with $m_t, m_s < 0$, switching
on the staggered magnetic field and
assuming that
$|m_s| \gg |m_t|$, we can integrate the singlet mode out to arrive again at the effective
model (\ref{eff.trip.ham}), with the renormalized masses still given by 
Eqs.(\ref{renorm.masses}). The important difference with the previous case
of the standard ladder is that the doublet mass gap now
increases with $h$ while the mass $m_3$ decreases and vanishes at the same
critical value as before. But since this time only one
Majorana mode becomes massless, the criticality is of the Ising type.
The dimerization order parameter, which is nonzero at $h < h_c$, vanishes at
the critical point as 
$$
\la \eps^- \ra \sim (h_c - h)^{1/8} \theta(h_c - h).
$$
The  (static) staggered magnetic susceptibility $\chi_{\rm stag} (h)$ is
constant in the zero-field limit,
\be
\chi_{\rm stag} (0) \sim (1/\alpha) (|m_t|/|m_s|)^{1/4}(|m_t|+|m_s|)^{-1},
\ee
then increases with $h$ and becomes logarithmically divergent at the
transition 
\be
\chi_{\rm stag} (h) \sim \ln (|h-h_c|/h_c).
\ee

Estimation of the large-distance asymptotics of the correlation functions 
is similar to what has been done in section 5. 
One only has to keep in mind that the only mass which changes its sign
across the transition is $m_3$.
Here we present the final results.

\medskip

In the dimerized phase ($h<h_c$) the dynamical structure factor
is entirely incoherent both in its longitudinal and transverse
components:
\bea
S^{+-} (\omega,\pi+q,\pi) &\propto& 
\left( \frac{l_3}{l_s} \right)^{{1}/{4}}\nn\\
&\times&
\frac{\theta[s^2 - (|m_d| + |m_3|)^2]}
{\sqrt{s^2 - (|m_d| + |m_3|)^2}},\nn\\
\label{chi-perp-gen}\\
S^{zz} (\omega,\pi+q,\pi) &\propto& 
\left( \frac{l_d^2}{l_sl_3} \right)^{\frac{1}{4}}
\frac{\theta[s^2 - 4m^2 _d]}
{\sqrt{s^2 - 4m^2 _d}}.
\label{chi-par-gen}
\eea
At the Ising criticality
\bea
S^{+-} (\omega,\pi+q,\pi) &\propto& 
\left( \frac{v^2}{l_sl_d} \right)^{\frac{1}{4}}
\frac{\theta[\omega^2 - q^2 v^2 - m_d ^2]}
{\left[s^2 - m_d ^2 \right]^{{3}/{4}}},\nn\\
\label{chi-perp-crit-gen}\\
S^{zz} (\omega,\pi+q,\pi) &\propto&
\left( \frac{l_d^3}{v^2l_s} \right)^{\frac{1}{4}}
\frac{\theta[s^2 -4 m^2 _d]}
{\left[s^2 - 4 m^2 _d \right]^{\frac{1}{4}}}.\nn\\
\label{chi-par-crit-gen}
\eea
In the region $h > h_c$, the longitudinal staggered spin fluctuations
remain incoherent:
\bea
&&S^{zz} (\omega,\pi+q,\pi) \propto
\left(\frac{l_d^2l_3(2l_2+l_3)^2}{l_sv^4}\right)^\frac{1}{4}\nn\\
&&\times\quad \theta[s^2 - (2|m_d| + m_3)^2],
\label{chi-par->}
\eea
whereas the transverse part
of the dynamical structure factor displays a coherent
$\delta$-function peak, 
\bea
S^{+-}(\omega,\pi+q,\pi) &\propto&
\left(\frac{v^4}{l_d^2l_sl_3}\right)^{\frac{1}{4}}\
\delta \left(s^2 - m^2 _d \right),
\label{chi-perp->}
\eea
describing a massive magnon with the spin projection $|S^z| = 1$. 
As in the standard ladder,  
at $h>h_c$ the dynamics of $\eps_-$ also displays a coherent mode
with the mass $|m_3|$.

Thus, characterization of the massive phase occuring at $h > h_c$ in the
spontaneously dimerized  ladder coincides with that for the standard ladder. This phase
occupies the region separated by the U(1) and Ising critical surfaces
on Fig.\ref{fig:fig1}.

\section{String order parameter}

Den Nijs and Rommelse \cite{nr} and Girvin and Arovas \cite{ga} have shown that
the Haldane gapped phase of the spin-1 chain is characterized by a nonlocal 
topological string order parameter,
\[
\la {\cal O}^{\alpha} \ra =
\lim_{|n-m|\to \infty} \la S^{\alpha}_n \exp ( \ri \pi
\sum_{j=n+1}^{m-1}S^{\alpha}_j ) S^{\alpha}_m
\ra, ~~(\alpha = x,y,z),
\]
whose nonzero value is associated with the breakdown of a hidden $Z_2 \times Z_2$
symmetry \cite{kt}.  For a weakly coupled SU(2)-symmetric spin-1/2 Heisenberg ladder,
the string order parameter, defined as
\be
{\cal O}^{\alpha}_{n,m} =
\prod_{j=n}^{m} \left( - 4 S^{\alpha}_{1j} S^{\alpha}_{2j} \right)
= \exp \left[ \ri \pi \sum_{j=n}^m (S^{\alpha}_{1j} + S^{\alpha}_{2j}) \right],
\label{string-op}
\ee
was discussed in Refs.\cite{snt,nt} (see also \cite{gnt}).
The description of the low-energy degrees of freedom of the spin ladder
in terms of
the Ising variables is especially efficient in this case
because the operator (\ref{string-op}) acquires a simple {\sl local} form in
terms of the Ising operators $\s_{\alpha}, \mu_{\alpha}$
$(\alpha=1,2,3)$, and the hidden $Z_2 \times Z_2$
symmetry becomes manifest. In the continuum limit\cite{comment}
\bea
\lim_{|x-y| \to \infty}\la {\cal O}_{\alpha} (x,y) \ra
&\equiv& \la {\cal O}_{\alpha} \ra
\sim \la \s_{\beta} \s_{\gamma} \ra^2
+ \la \mu_{\beta}  \mu_{\gamma} \ra^2,\nn\\
&&\alpha \neq \beta\neq \gamma\ ,
\label{string-op-cont}
\eea
so that the string order parameter reveals the SU(2) symmetry and
is nonzero both in the Haldane 
and dimerized 
phases just because the degenerate triplet of the Ising systems is either 
disordered ($\la \s_{\alpha} \ra =0$, $\la \mu_{\alpha} \ra \neq 0$)
or ordered ($\la \s_{\alpha} \ra \neq 0$, $\la \mu_{\alpha} \ra = 0$).

As we have seen, the staggered magnetic field removes the SO(3) degeneracy
of the triplet modes, and in the massive phase located between the two critical
surfaces shown in Fig.\ref{fig:fig1} (the case $h > h_c$), 
the signs of the masses $m_1 = m_2$ and $m_3$ are opposite. 
As a result, the dependence of
the longitudinal and transverse components of the string order parameter
on $h$ becomes qualitatively different.

\subsection{U(1) transition in the Haldane phase}

Let us start with the longitudinal string order parameter, ${\cal O}_z$. 
In the region of small fields, $0 < h \ll h_c$, one can adopt the picture
of three independent triplet Ising copies as the zero-order approximation
and take into account the effect of the staggered field as a small perturbation.
According to (\ref{renorm.masses}) and (\ref{renorm.couplings}), the latter
leads to splitting of the triplet masses and renormalization of the coupling
constants, both being of the order of $h^2$. As a result
\bea
\la  {\cal O}_z \ra (h) &\simeq&  \la \mu_1 \ra \la \mu_2 \ra\nn\\
&=& \la  {\cal O}_z \ra (0)\left[
1 - \frac{C_1}{2} \left( \frac{l_t}{\alpha} \right) 
\left(\frac{\bar{h}}{m_s}\right)^2 + O (\bar{h}^4) \right],\nn\\
\label{Oz-small-h}
\eea 
where $\la  {\cal O}_z \ra (0) \sim (m_t \alpha/v)^{1/2}$.
In vicinity of the critical point,
$|h - h_c| \ll h_c$, the Ising doublet (1,2) becomes very soft and asymptotically
decouples from the rest of the spectrum, being described
in terms of the SGM (\ref{sg}). In this case the operator
${\cal O}_z$ can be bosonized 
(see Appendix A).
However, since $K \neq 1$, in
formulas (\ref{br3}) and (\ref{br4}) for the products
of Ising operators
$\mu_1 \mu_2$ and $\s_1\s_2$, one 
should rescale the field $\Phi$: 
$\Phi \to \sqrt{K} \Phi$. So
\be
\la  {\cal O}_z \ra \sim \la \sin \sqrt{\pi K} \Phi \ra^2 +
\la \cos \sqrt{\pi K} \Phi \ra^2.
\label{O-z1}
\ee
Since $m_d \sim h_c - h$, at $h < h_c$ $\la \sin \sqrt{\pi K} \Phi \ra = 0$, and
\be
\la  {\cal O}_z \ra \sim \la \cos \sqrt{\pi K} \Phi \ra^2 \sim M_d ^{K/2}
\sim (h_c - h)^{\frac{K}{4-2K}}. \label{Oz<}
\ee
At $h>h_c$ $\la \cos \sqrt{\pi K} \Phi \ra = 0$, and
\be
\la  {\cal O}_z \ra \sim \la \sin \sqrt{\pi K} \Phi \ra^2 \sim |M|_d ^{K/2}
\sim (h - h_c)^{\frac{K}{4-2K}}. \label{Oz>}
\ee 
Formulas (\ref{Oz<}), (\ref{Oz>}) determine the power law according to which
the string order parameter $\la  {\cal O}_z \ra$ vanishes at the transition.

Upon increasing the field in the region $h > h_c$ $\la  {\cal O}_z \ra$ keeps growing.
In the limit of strong fields, $h \gg h_c$, the system represents two identical copies
of the Heisenberg S=1/2 chain in a staggered magnetic field, each of them being
represented in the continuum limit by a $\beta^2 = 2\pi$ SGM with the nonlinear term
proportional to $h \cos \sqrt{2\pi}\Phi_a~(a=1,2)$. Clearly, the field
$\Phi$ is a symmetric linear combination of $\Phi_1$ and $\Phi_2$:
\[
\Phi = \frac{\Phi_1 + \Phi_2}{\sqrt{2}}.
\]
Therefore, in this limit
\bea
\la  {\cal O}_z \ra &\sim& \la \cos \sqrt{\frac{\pi}{2}} \Phi_1 \ra^2
\la \cos \sqrt{\frac{\pi}{2}} \Phi_2 \ra^2\nn\\
&+& \la \sin \sqrt{\frac{\pi}{2}} \Phi_1 \ra^2
\la \sin \sqrt{\frac{\pi}{2}} \Phi_2 \ra^2 \nn\\
&+& \la \cos \sqrt{\frac{\pi}{2}} \Phi_1 \ra^2
\la \sin \sqrt{\frac{\pi}{2}} \Phi_2 \ra^2\nn\\
&+& \la \sin \sqrt{\frac{\pi}{2}} \Phi_1 \ra^2
\la \cos \sqrt{\frac{\pi}{2}} \Phi_2 \ra^2. \label{Oz3}
\eea
The minima of the potentials $h \cos \sqrt{2\pi}\Phi_a$ are
$\left( \Phi_a \right)_m = \sqrt{2\pi} (m + 1/2)$ at $h>0$ and
$\left( \Phi_a \right)_m = \sqrt{2\pi} m$ at $h < 0$, where $m=0,\pm 1,\pm 2, ...$.
Therefore, the last two terms in (\ref{Oz3}) vanish, and for any sign of $h$
the $z$-component of the string order parameter grows as
\be
\la  {\cal O}_z \ra \sim |h|^{1/3}.
\label{Oz4}
\ee

Consider now the transverse components of the string order parameter, 
$\la  {\cal O}_x \ra = \la  {\cal O}_y \ra$. 
At $h \ll h_c$ the behaviour of $\la  {\cal O}_x \ra$ is similar to that for
$\la  {\cal O}_z \ra$:
\bea
\la  {\cal O}_x \ra (h) &\simeq&  \la \mu_2 \ra^2 \la \mu_3 \ra^2\nn\\ 
&=& \la  {\cal O}_z \ra (0)
\Bigl[1 - \frac{1}{4} \left( \frac{l_s}{\alpha} \right)
\left(\frac{C_2}{\pi} + C_1 \frac{g_1}{\pi v} \frac{|m_s|}{m_t}
\right)\nn\\
&&\qquad\times \left(\frac{\bar{h}}{m_s}\right)^2\ln\left(\frac{l_t}{\alpha}\right) 
+ O (\bar{h}^4) \Bigr].
\label{Ox-small-h}
\eea 
Near the critical field, where the doublet of the Ising systems decouples from the
third Ising component of the triplet, one finds that
\bea
\la {\cal O}_x \ra (h) &\sim& \left( \frac{M_d \alpha}{v} \right)^{1/4}
\theta(h_c - h) \nn\\
&\sim& (h_c - h)^{\frac{1}{4(2-K)}} \theta(h_c - h).\label{Ox-crit}
\eea
Thus, due to the fact that at $h > h_c$ the Ising doublet becomes ordered
($\la \mu_1 \ra = \la \mu_2 \ra = 0$), in the large-$h$ massive phase 
the transverse components of the string order parameter vanish.
\begin{figure}[ht]
\begin{center}
\epsfxsize=0.4\textwidth
\epsfbox{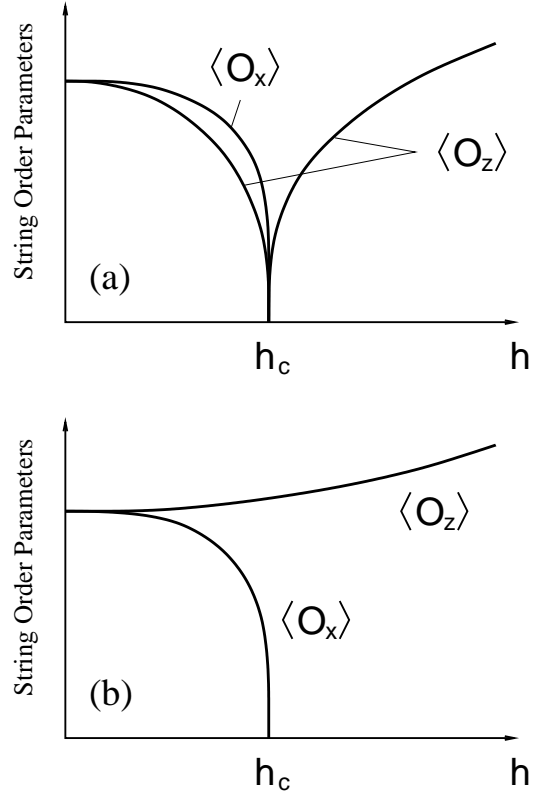}
\end{center}
\caption{String order parameters as functions of the staggered magnetic 
field across (a) the U(1) transition in the Haldane phase and (b) 
the Ising transition in the dimerized phase.} 
\label{fig:sop1}
\end{figure}

\subsection{Ising transition in the dimerized phase}

In the spontaneously dimerized, non-Haldane phase the Ising doublet remains massive
at any $h$. The mass gap 
$|m_d|$ is an increasing function of $h$; so is the longitudinal string order parameter.
At $h \ll h_c$
\bea
\la  {\cal O}_z \ra (h) &\simeq&  \la \s_1 \ra \la \s_2 \ra \nn\\
&=& \la  {\cal O}_z \ra (0)\left[
1 + \frac{1}{2} C_1 \left( \frac{l_t}{\alpha} \right) 
\left(\frac{\bar{h}}{m_s}\right)^2 + O (\bar{h}^4) \right].\nn\\
\label{Oz-small-h-dim}
\eea
$\la  {\cal O}_z \ra$ grows monotonically with $h$ and at large $h$ ($h \gg h_c$) crosses
over to the $|h|^{1/3}$ behaviour.
At $h \ll h_c$ $\la  {\cal O}_x \ra$ is still given by formula (\ref{Ox-small-h}), whereas
close to the Ising criticality
\be
\la  {\cal O}_x \ra \sim (h_c - h)^{1/4} \theta (h_c - h).
\label{Ox-crit-dim}
\ee
The dependence of the string order parameters on the staggered field is 
schematically shown shown in Figs.\ref{fig:sop1}.


\section{Conclusions}


In this paper, we have analyzed the properties of the two-leg
antiferromagnetic spin-1/2 ladder in a staggered magnetic field. We
have considered the spin-liquid phase of the standard ladder and the spontaneously
dimerized phase of a generalized spin-ladder model. We have shown that in the
former case the staggered field drives the system towards a Gaussian criticality,
whereas in the latter case it induces an Ising transition. These two criticalities
are associated with a softening of the transverse ($S^z = \pm 1$)
and longitudinal ($S^z = 0$) collective modes, respectively, and
are characterized by power-law and logarithmic divergencies of the 
staggered magnetic susceptibility.

By comparing our results for weakly ($J_{\perp}/J \ll 1$) and
strongly coupled ($J_{\perp}/J \gg 1$) ladders in a staggered field,
we can identify certain universal features of the transition. The very
existence of the Gaussian criticality is universal. At criticality the
low-energy degrees of freedom can be decribed in terms of spin-1/2
{\sl spinons} similar to those found in the anisotropic spin-1/2 Heisenberg
XXZ chain. This is interesting and shows that the staggered field $h$
leads to a destabilisation of the magnons that form the low-lying part
of the spectrum in the absence of $h$ and eventually ``deconfines''
them into pairs of spinons.

Close to this criticality the low-energy part of the
spectrum involves a doublet of interacting transverse
modes. These are described in terms of an effective spin-1/2 XXZ chain
in a weak staggered field, which vanishes at $h=h_c$. The transverse
modes are identified with {\sl quantum solitons} (with gap $M_d$) 
of an underlying SGM. These solitons together with
soliton-antisoliton bound states determine the behaviour of the
dynamical structure factor in both massive phases ($h<h_c$ and
$h>h_c$). The soliton reveals itself as a {\sl coherent} delta-function
peak in the transverse staggered structure factor $S^{\pm}(\omega, \pi
+ q, q_{\perp} = \pi)$. 

On the other hand, the $q_\perp=0$ part of the spin excitation spectrum
is different in the weak and strong coupling cases. The same hold true
for the $q_\perp=\pi$ part of the longitudinal structure factor: at
weak coupling the $S^z=0$ mode is still seen in the low-energy part of
the spectrum, whereas it is pushed to very high energies in the
strong coupling regime.

In the absence of a staggered field the generalised ladder is
spontaneously dimerized and its elementary excitations can be
understood in terms of topological dimerization kinks \cite{nt}.
When a staggered field is applied the dimerization diminishes but
the qualitative picture remains unchanged until the field reaches its
critical value $h_c$, where we find an Ising criticality. Here once
again we may think of the elementary excitations in terms of pairs of
spinons. The physical properties for $h>h_c$ are the same as in the
$h>h_c$ phase of the standard ladder discussed above (see also
Fig.\ref{fig:fig1}). In particular, the coherent $S^z=\pm 1$ modes are
recovered. This identification is confirmed by the behaviour of the
longitudinal and transverse string order parameters.

As we have seen, the physics of the spin-1/2 ladder in a staggered
field is very rich. We think that it would be very interesting to
explore it experimentally. 
An open question is to analyze the crossover region between weak and
strong coupling regimes by e.g. numerical methods.

\bigskip

\acknowledgements
\medskip

The authors express their sincere gratitude to A.M. Tsvelik and
R. Coldea for helpful discussions. AAN thanks the Department of
Physics at Warwick University for hospitality. 
This work was supported by the EPSRC under grants AF/100201 (FHLE) and
GR/N19359 (FHLE and AAN) and by the INTAS-Georgia grant No. 97-1340
(AAN). 

\appendix

\section{Some facts about 2D Ising model}\label{ising}

In this Appendix we briefly summarize those facts about the 2D Ising model which
are used in the main part of the paper and other Appendices. 

Close to criticality ($|T-T_c|\ll T_c$), the  
scaling properties of the 2D
Ising model are described by a Lorentz-invariant
(1+1)-dimensional quantum model of a massive
real (Majorana) fermion\cite{zi,difran}. The 
corresponding 2D Euclidean action (written in complex notations:
$z=\tau+\ri x$, $\bz=\tau-\ri x$, $\p=\p/\p z$, $\bp=\p/\p \bz$; $v=1$)
reads\cite{prefactors}:
\be
S = \int \rd^2 z~ \left( \xi_L \bp \xi_L + \xi_R \p \xi_R + \ri m \xi_L \xi_R \right).
\label{ising-maj-action}
\ee
Here 
$\xi_L$ and $\xi_R$ are the holomorphic (left) and antiholomorphic
(right) components of the fermionic field.
The magnitude of the  mass,
\[
m \sim \left(\frac{v}{\alpha}\right)\left(\frac{T-T_c}{T_c} \right).
\]
($\alpha$ being a short-distance cutoff)
determines the correlation length in the Ising model,
$l_c \sim v/|m| \gg \alpha$, 
which diverges at criticality ($m=0$),
and the sign of the mass indicates whether the system is
ordered ($m < 0$) or disordered ($m>0$). 
The set of strongly fluctuating fields of the Ising model 
(at criticality these are known as primary fields of the conformal field theory
with central charge $C=1/2$)
includes the fermion
field ($\xi_L,\xi_R$), the mass bilinear (or energy density) $\vare = \ri \xi_R \xi_L$,
and order and disorder fields, $\s$ and $\mu$. The latter two fields are
nonlocal with respect to each other and also with respect to the Majorana field.

Two identical noninteracting copies of the 2D Ising models are described
by a pair of Majorana fermions, $\xi^1$ and $\xi^2$, which can be combined
into a complex (Dirac) massive fermion, $\psi = (\xi^1 + \ri \xi^2)/\sqrt{2}$.
The latter can be bosonized\cite{zi}. If the two Ising copies are slightly
noncritical, the resulting bosonic theory represents a quantum sine-Gordon
model at the decoupling (or Luther-Emery) point $\beta^2 = 4\pi$: 
\be
S_{SG} = \int \rd^2 z~ \left[ \frac{1}{2} (\p_{\mu} \Phi)^2
- \frac{m}{\pi\alpha} \cos \sqrt{4\pi} \Phi \right].
\label{SG-4pi}
\ee
Below we give a list of bosonization 
rules for two Ising copies\cite{difran,gnt,fgn} which are used in the main text:
\begin{itemize}
\item{} The chiral components of the U(1) current:
\bea
&&J_L (z)= \ri \xi^1 _L (z)\xi^2 _L (z) = \frac{\ri}{\sqrt{\pi}} \p \phi_L (z),\nn\\
&&J_R (\bz) = \ri \xi^1 _R (\bz)\xi^2 _R (\bz)= - \frac{\ri}{\sqrt{\pi}} \bp \phi_R (\bz). \label{br1}
\eea
\item{} The total energy density:
\be
\vare_1 (z,\bz)+ \vare_2 (z,\bz)=  \frac{1}{\pi\alpha}
\cos \sqrt{4\pi} \Phi (z,\bz). \label{br2}
\ee
\item{} The fermionic fields:
\bea
&&\xi^1 _L (z)+ \ri \xi^2 _L (z)\simeq (\pi\alpha)^{-1/2} \exp [-\ri \sqrt{4\pi} \phi_L (z)],\!
\label{fer-bosL}\\
&&\xi^1 _R (\bz)+ \ri \xi^2 _R (\bz)\simeq (\pi\alpha)^{-1/2} \exp [\ri \sqrt{4\pi} \phi_R (\bz)].\!
\label{fer-bosR}
\eea
\item{}
Mixed products of the order and disorderd operators (a more accurate definition
of these products includes Klein factors\cite{fgn}):
\bea
&& \s_1 \s_2 \sim \sin \sqrt{\pi} \Phi, ~~\mu_1 \mu_2 \sim \cos \sqrt{\pi} \Phi,
 \label{br3}\\
&& \s_1 \mu_2 \sim \cos \sqrt{\pi} \Theta, ~~\mu_1 \s_2 \sim \sin \sqrt{\pi} \Theta.
 \label{br4}
\eea
\end{itemize}
In the above formulas,
\[
\Phi(z,\bz) = \phi_L(z) + {\phi}_R(\bz)
\]
in the scalar field of the underlying SGM, and
\[
\Theta(z,\bz) = \phi_L(z) - {\phi}_R(\bz).
\]
is its dual counterpart.

Using these bosonization rules, one can easily recover all OPEs for a single Ising model.
In particular, fusing the products of Ising operators in (\ref{br3}) one derives the
OPEs 
\bea
&&\s(z,\bz) \s(w,\bw) \nn\\
&& \sim \frac{1}{\sqrt{2}} \left( \frac{\alpha}{|z-w|} \right)^{1/4}
\left[ 1 + \pi |z-w| \vare (w,\bw)\right], \label{ope11} \\
&&\mu (z,\bz) \mu (w,\bw) \nn\\
&& \sim \frac{1}{\sqrt{2}} \left( \frac{\alpha}{|z-w|} \right)^{1/4}
\left[ 1 - \pi |z-w| \vare (w,\bw)\right], \label{ope12}
\eea
used in section II B. In the same way, from 
representation (\ref{br4}) one derives two more OPEs:
\bea
&&\s(z,\bz) \mu(w,\bw) \nn\\
&&\sim\sqrt{\frac{\pi}{2}}\frac{\gamma (z-w)^{1/2} \xi_L(w) + \gamma^* (\bz-\bw)^{1/2} \xi_R(\bw)}
{(|z-w|/\alpha)^{1/4}},\label{OPE5}\\
&&\mu(z,\bz) \s(w,\bw) \nn\\
&&\sim\sqrt{\frac{\pi}{2}}\frac{\gamma^* (z-w)^{1/2} \xi_L(w) + \gamma (\bz-\bw)^{1/2} \xi_R(\bw)}
{(|z-w|/\alpha)^{1/4}},\label{OPE6}
\eea
where $\gamma = e^{\ri \pi/4}$.

Since the SGM (\ref{SG-4pi}) occurs in a topologically ordered, massive phase,
in the ground state the field $\Phi$ is locked in one of the infinitely degenerate minima of
the potential ${\cal U} = - m \cos \sqrt{4\pi} \Phi$:
\bea
\left( \Phi \right)^{\rm vac}_n = \sqrt{\pi} n, &&~{\rm if}~ m > 0, \nn\\
\left( \Phi \right)^{\rm vac}_n = \sqrt{\pi} (n + 1/2),
&&~{\rm if}~ m < 0.
\eea 
From (\ref{br3}) it then follows that
\bea
\la \s_{1,2}\ra = 0, ~~\la \mu_{1,2} \ra \neq 0, &&~{\rm if}~ m > 0, \nn\\
\la \s_{1,2}\ra \neq 0, ~~\la \mu_{1,2} \ra = 0, &&~{\rm if}~ m < 0.
\label{phases}
\eea
Quantum solitons of the SGM (\ref{SG-4pi}) are associated with
the vacuum-vacuum transitions, $\Phi \to \Phi \pm \sqrt{\pi}$,
$\Theta\to\Theta$, which correspond to the following
Z$_2$$\times$Z$_2$ transformations of the fermionic fields and Ising
operators: 
\bea
&&\xi^a_L \to - \xi^a_L,~~\xi_R^a \to - \xi_R^a ~~(a=1,2), \nn\\
&&\s_1 \to \mp \s_1, ~\s_2 \to \pm \s_2, ~ \mu_1 \to \pm \mu_1,~\mu_2 \to \mp \mu_2 \label{trnsf}
\eea
For any sign of $m$, this symmetry is spontaneously broken
in one Ising copy and preserved in the other,
leading to the conclusion that quantum solitons of the model (\ref{SG-4pi})
describe kinks of a single ordered (disordered) Ising system that connect
opposite values of the order (disorder) parameters.

Parity (or charge conjugation) transformations 
\bea
m > 0:&& ~~\xi^1_{R,L}\to \xi^1_{R,L},~~~ \xi^2_{R,L}\to -\xi^2_{R,L},\nn\\
&& ~~\Phi \to - \Phi,~~~\Theta \to - \Theta, \nn\\
&& ~~ \s_1 \s_2 \to - \s_1 \s_2, ~~\mu_1 \mu_2 \to \mu_1 \mu_2 ,\nn\\
&&~~\s_1 \mu_2 \to \s_1 \mu_2 , ~~\mu_1 \s_2 \to - \mu_1 \s_2 
\label{parity1}
\eea
and
\bea
m < 0: && ~~\xi^1_{R,L}\to \xi^1_{R,L},~~~ \xi^2_{R,L}\to -\xi^2_{R,L},\nn\\
&& ~~\Phi \to \sqrt{\pi} - \Phi,~~~\Theta \to \sqrt{\pi} - \Theta, \nn\\
&& ~~ \s_1 \s_2 \to  \s_1 \s_2, ~~\mu_1 \mu_2 \to - \mu_1 \mu_2 ,\nn\\
&&~~\s_1 \mu_2 \to - \s_1 \mu_2 , ~~\mu_1 \s_2 \to \mu_1 \s_2 
\label{parity2}
\eea
keep invariant vacuum expectation values of the order (disorder) parameters at $m<0$ ($m>0$)
and therefore serve as a tool to conclude whether a given correlation function in a
broken-symmetry phase is nonzero.
For example, consider the correlation function
\be
K ({\bf r}) = \la \mu ({\bf r}) \vare ({\bf 0})\ra
\ee
for a single {\sl ordered} Ising model. For two identical and decoupled Ising models
this correlator can be squared:
\be
K_2 ({\bf r}) = \la \mu_1 ({\bf r}) \mu_2 ({\bf r}) \vare_1 ({\bf 0})\vare_2 ({\bf 0})\ra
= K^2 _0 ({\bf r}). 
\ee
Under transformations (\ref{parity2}) the product $\vare_{1}\vare_{2}$ stays intact but
$\mu_1 \mu_2$ changes its sign. Therefore
\be
K_2 ({\bf r}) = K^2 _0 ({\bf r}) = 0.
\label{zero}
\ee
This fact has been used in section IIB.

\section{Field induced admixture between the singlet and triplet modes}
\label{admix}

In this Appendix we consider two, apparently ``high-energy'', operators:
the total staggered magnetization, ${\bf n}^+$, and the smooth part of the relative magnetization, 
${\bf K}$, defined in (\ref{n+}) and (\ref{vec-currents-maj}),
and find their projections onto the low-energy, triplet sector of the model.

\subsection{Projecting ${\bf n}^{\bf +}$}

As follows from the comparison of (\ref{op-transmut})
with formula (\ref{eff.action2}),
the low-energy projection of the operator
${\cal O}_0 = n^+ _z$ has actually been found in section IIB, and the result is
contained in the second-order correction to the effective action (\ref{eff.trip.action}).
Thus we arrive at Eq.(\ref{proj-n+z}).

Consider now the operator ${\cal O}_0 = n^+ _x$. Treating all Ising systems as
decoupled, we have: 
\bea
&&\la n^+ _x ({\bf r}) n^+ _z ({\bf r}_1) \ra_s 
 = \alpha^{-2} [\mu_1 ({\bf r}) \s_1 ({\bf r}_1)][ \s_2 ({\bf r}_1) \s_2 ({\bf r}_1)]\nn\\
&& \times
[\s_3 ({\bf r}) \mu_3 ({\bf r}_1)] \la \mu_4 ({\bf r}) \mu_4 ({\bf r}_1)\ra_s \nn 
\eea
The correlator $\la \mu_4 ({\bf r}) \mu_4 ({\bf r}_1)\ra_s$ is
short-ranged (see (\ref{mu4-mu4-corr})). Hence the products of
operators in the square brackets, all of them defined in the triplet
sector, are subject to fusion.  Using the fusion rules (\ref{ope11}),
(\ref{OPE5}) and (\ref{OPE6})  and integrating over the relative coordinate
$\vrho = {\bf r} - {\bf r}_1$, we arrive at the result (\ref{proj-n+x}).
In the same way one obtains the low-energy projection for $n^+ _y$ given by (\ref{proj-n+y}).

\subsection{Projecting K}

Here we derive the low-energy projection of the operator
\[
{\bf K} = \ri (\vxi_R \xi^4 _R + \vxi_L \xi^4 _L ).
\]
Consider first the case
${\cal O}_0 = K_x$. The correlator in (\ref{op-transmut}) reads:
\bea
\la K_x ({\bf r}) n^+ _z ({\bf r}_1) \ra_s &&= (\ri/\alpha)
\s_2 ({\bf r}_1)\mu_3 ({\bf r}_1)\nn\\
&&\times \sum_{\nu=R,L}\xi^1 _{\nu} ({\bf r}) \s_1 ({\bf r}_1)\la \xi^4 _{\nu} (\bf {\bf r})
\mu_4 ({\bf r}_1)\ra_s
\eea
Keeping in mind that $m_s < 0$, first we note that the correlators
$\la \xi^4 _{\nu} (\bf {\bf r})\mu_4 ({\bf r}_1)\ra_s$ are invariant under charge conjugation
(\ref{parity2}) and, therefore, are nonzero.
 These correlators are chiral
but otherwise short-ranged, decaying exponentially at $|{\bf r} - {\bf r}_1| \sim l_s$.
Since they serve as integral kernels, a qualitatively correct
estimation of the integral in (\ref{op-transmut}) can be obtained 
if one treats the product
\be
\left[ \xi^1 _L (z) \xi^4 _L (z) + \xi^1 _R (\bz) \xi^4 _R (\bz)\right]
\s_1 (z_1,\bz_1) \mu_4 (z_1,\bz_1)
\label{prodK}
\ee
by OPE and then confines the integration region to the interval
$0 < |{\bf r} - {\bf r}_1| \leq l_s$.
To proceed further, one can 
bosonize two local products, $\xi^1 \xi^4$
and $\s_1 \mu_4$, and then fuse these fields as those belonging to
a (critical) Gaussian model. Since we are looking for a short-distance
OPE, the relative sign of the Majorana masses $m_1$ and $m_4$ is unimportant,
and bosonization rules (\ref{br3}), (\ref{br4}) are perfectly applicable. 
We obtain:
\bea
&&\frac{1}{\sqrt{\pi}} \left[\p \phi_L(z) - \bp \phi_R (\bz)  \right]
\cos \sqrt{\pi} \Theta (z_1,\bz_1) \nn\\
&& ~~\sim \frac{1}{4\pi} \left( \frac{1}{z-z_1} + \frac{1}{\bz - \bz_1} \right)
\sin \sqrt{\pi} \Phi (z_1,\bz_1) \nn\\
&& ~~ =  \frac{1}{2\pi} \frac{\Re e (z-z_1)}{|z-z_1|^2} \mu_1 (z_1,\bz_1) \s_4 (z,\bz_1),
\label{14-ope}
\eea
and therefore
\be
\la K_x ({\bf r}) n^+ _z ({\bf r}_1) \ra_s \simeq 
 \frac{\ri}{2\pi}  \frac{\tau - \tau_1}{|{\bf r} - {\bf r}_1|^2} 
N^- _y ({\bf r}_1),
\label{result-int}
\ee
where 
$
N^- _y  \sim \alpha^{-1}\mu_1 \s_2 \mu_3 \la \s_4 \ra
$
is the $y$-component of relative staggered magnetization averaged over the high-energy singlet modes.
According to the definition (\ref{op-transmut}) 
\[
\delta  K_x ({\bf r}) =   \frac{\ri h}{2\pi v}
\int_{\rho < \xi_s} \rd^2 \vrho \left(\frac{\rho_0}{\rho^2}\right) N^- _y ({\bf r} - {\vrho}),
\]
where
\[
\vrho = {\bf r} - {\bf r}_1 = (\rho_0, \rho_1) =
(v\tau, x).
\]
Expanding in $\rho$, the lowest-order projection of $K_x$ onto the triplet sector
is found to be given by formula (\ref{proj-Kx}).
Quite similarly one arrives at formula (\ref{proj-Ky}) for the projection of $K_y$  .

In the case ${\cal O}_0 = K_z$, we follow the same procedure and use (\ref{br3})
to bosonize the product $\mu_3 \mu_4$. The corresponding OPE reads:
\bea
&&\left[ \xi^1 _L (z) \xi^4 _L (z) + \xi^1 _R (\bz) \xi^4 _R (\bz)\right]
\mu_3 (z_1,\bz_1) \mu_4 (z_1,\bz_1)\nn\\
&& \sim \frac{1}{\sqrt{\pi}} \left[\p \phi(z) - \bp \bar{\phi} (\bz)  \right]
\cos \sqrt{\pi} \Phi (z_1,\bz_1) \nn\\
&& ~~=  \frac{1}{4\pi} \left( \frac{1}{z-z_1} - \frac{1}{\bz - \bz_1} \right)
\sin \sqrt{\pi} \Phi (z_1,\bz_1) \nn\\
&& ~~ =  - \frac{\ri}{2\pi} \frac{\Im m (z-z_1)}{|z-z_1|^2} \s_3 (z_1,\bz_1) \s_4 (z,\bz_1),
\nn
\eea
and this eventually leads to formula (\ref{proj-Kz}) for the low-energy projection
of $K_z$.

\section{Energy-density correlations in the doublet sector}\label{eps-corr}

\begin{figure}[ht]
\begin{center}
\epsfxsize=0.25\textwidth
\epsfbox{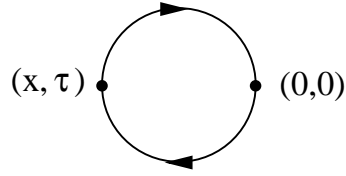}
\end{center}
\caption{Two-fermion bubble.} 
\label{fig:bubble}
\end{figure}

Here we estimate the Fourier transform of the 2-point energy-density correlation 
function in the doublet sector,
\bea
&&K_d (q,\omega_m) = \int \rd x \rd \tau K_d (x,\tau)
e^{- \ri (qx - \vare \tau)}, \label{K_d} \\
&&K_d (x,\tau) = \la T_{\tau} \vare_d (x,\tau) \vare_d (0,0) \ra, 
~~
\vare_d = \ri\sum_{a=1,2} \xi^a _R \xi^a _L, \nn
\eea
and find its analytical continuation ($\ri \omega_n \to \omega + \ri \delta$)
that determines the structure factor $S^{zz} (\omega, \pi+q, 0)$. We
will consider the case $m_d \neq 0$ and assume that the two massive
Majorana fermions are decoupled.  
In this case
$
K_d ({\bf r}) = 2~ {\rm det} \hat{G} ({\bf r}), 
$
where $\hat{G} ({\bf r})$ is the real-space 2$\times$2 Green's function matrix for a free
massive fermion in the Nambu representation. We have (see Fig. \ref{fig:bubble})
\bea
&&K_d (q,\omega_n) = 2 \int \frac{\rd k}{2\pi} \frac{\rd \vare}{2\pi}
[~G_{RR} (k_+,\vare_+) G_{LL} (- k_-, - \vare_-)\nn\\
&&~~ -
 G_{RL} (k_+,\vare_+) G_{LR} (- k_-, - \vare_-) ~],
\label{K_d-fourier}
\eea
where
\bea
\hat{G} (k,\vare) &=& \left(
\begin{array}{clcr}
G_{RR} (k,\vare),&G_{RL} (k,\vare)\\
G_{LR} (k,\vare),&G_{LL} (k,\vare)
\end{array}
\right) \nn\\
&=& - \frac{\ri \vare + kv \tau_3 + m_d \tau_2}{\vare^2 + k^2 v^2 + m^2 _d},
\label{GF-mom.repres}
\eea 
and $k_{\pm} = k \pm q/2$, $\vare_{\pm} = \vare \pm \omega_n /2$.

Since the mass bilinear $\xi_R \xi_L$ is a Lorentz-invariant object,
$K_d (q,\omega_n)$ depends only on $\omega^2 _n + q^2$. This makes it sufficient
to estimate $K_d (0,\omega_n)$ in which case the calculations become especially simple.
Integrating over $\vare$ in (\ref{K_d-fourier}) yields:
\bea
&&K_d (0,\omega_n) =  - \int\frac{\rd k}{2\pi} \frac{1}{\ri \omega_n E_k}\nn\\
&&\times \left[ \frac{\ri \omega_n E_k- 2m^2 _d}{\ri \omega - 2E_k}
- \frac{\ri \omega_n E_k + 2m^2 _d}{\ri \omega + 2E_k} 
\right]
\eea
where $E_k = \sqrt{k^2 v^2 + m^2 _d}$.
Analytically continuing this expression, then
taking the imaginary part and finally replacing $\omega^2$ by $s^2 = \omega^2 - q^2 v^2$,
 we get ($\omega > 0$):
\be
\Im m K_d (q,\omega + \ri \delta) = \frac{1}{2v}\frac{\sqrt{s^2 - 4m^2 _d}}{s}. 
\label{ed-ed-imaginary}
\ee
This result has been used in Eq.(\ref{Szz-stag-0}).

\section{Transverse structure factor at small momentum} \label{app-transverse}

In this Appendix we provide some technical details concerning
the small-$q$ structure factor $S^{xx} (\omega,q,0)$ at the critical point.
We start from the real-space representation  
of the Matsubara polarization operator $X^{xx} (x,\tau)$
\bea
X^{xx} (x,\tau) &=& \la T_{\tau} I^1 (x,\tau) I^1 (0,0) \ra\nn\\
&=& Tr \left[ \hat{G} (\tau,x) \hat{\cal G} (\tau,x) \right].
\label{X-realspace}
\eea
Here $\hat{G}$ and $\hat{\cal G}$
are
is the Green's function matrices for the massive and massless Majorana
fields $\xi^3$ and $\xi^2$, respectively.
Due to the marginal interaction in the doublet $(\xi^1,\xi^2)$ sector,
$\hat{\cal G}$ has the structure of the single-particle Green's function
of a spinless Tomonaga-Luttinger liquid with the interaction constant $K$. 
Using the explicit expressions for these two Green's functions,
\bea
G_{RR}(z,\bz) &=& - \frac{m_3}{2\pi} \sqrt{\frac{z}{\bz}} K_1 (m_3|z|), \nn\\
G_{LL}(z,\bz) &=& - \frac{m_3}{2\pi} \sqrt{\frac{\bz}{z}} K_1 (m_3|z|),
\label{massive-GF}
\eea
\bea
{\cal G}_{RR} (z,\bz) &=& - \frac{1}{2\pi \alpha}\left(\frac{z}{\bz} \right)^{1/2}
\left( \frac{\alpha}{|z|} \right)^{\frac{1}{2}(K+1/K)},\nn\\
{\cal G}_{LL} (z,\bz) &=& - \frac{1}{2\pi \alpha}\left(\frac{\bz}{z} \right)^{1/2}
\left( \frac{\alpha}{|z|} \right)^{\frac{1}{2}(K+1/K)},\nn\\
{\cal G}_{RL} (z,\bz) &=& 0,
\label{TL-propagators} 
\eea
we obtain:
\bea
&&X(z,\bz)\nn\\
&& = \frac{m_3\alpha}{(2\pi\alpha)^2} 
\left( \frac{\alpha}{|z|}\right)^{\frac{1}{2}(K+1/K)}
\left( \frac{z}{\bz} + \frac{\bz}{z} \right) K_1 (m_3|z|).
\label{X1}
\eea
Introducing two-dimensional real vectors, 
${\bf q} = (\omega_n, qv)$ and ${\bf r} = (\tau,x/v)$ 
($\omega_n$ being the Matsubara frequency), we
pass to the Fourier transform $X^{xx}({\bf q})$ and, after angular 
integration, obtain:
\bea
&&X^{xx}({\bf q}) = - \frac{1}{\pi} (m_3\alpha)^{2\theta}
\left(\frac{\omega^2 _n - q^2}{{\bf q}^2} \right)\nn\\
&&~~~~\times
\int_{0} ^{\infty} \rd x~x^{-2\theta} K_1 (x) J_2 \left(\frac{|{\bf q}|}{m_3}x\right),
\label{X4}
\eea
where $J_2 (x)$ is the Bessel function, and
\be
2\theta = \frac{1}{2}\left(K+\frac{1}{K}\right)-1. 
\label{theta*}
\ee
Since the expressions for the Tomonaga-Luttinger propagators
(\ref{TL-propagators}) are asymptotic (i.e. valid at $|z|>\alpha$)
the lower integral cutoff must be finite, $\sim m_3 \alpha$. 
On the other hand, the integral in (\ref{X4}) is convergent
at $x\to 0$, if $\theta < 1$. As follows from (\ref{theta*}),
this condition is satisfied
not only in the weak-coupling case, where $K$ is very close to 1, but also 
in the strong-coupling regime where
$K = 3/4$ (in the latter case $\theta \approx 2\cdot 10^{-2}$). 
This justifies the replacement of the lower integral limit by 0,
in which case the result can be expressed
in terms of a hypergeometric function:
\bea
&&X^{xx}({\bf q}) = - \frac{1}{\pi} 
\frac{\Gamma(2 - \theta)\Gamma(1 - \theta)}{2^{1+2\theta}\Gamma(3)}
(m_3\alpha)^{2\theta}\nn\\
&&~~~~\times\left(\frac{\omega^2 _n - q^2}{m^2 _3} \right)
F \left(1 - \theta, 2 - \theta; 3; - \frac{{\bf q}^2}{m^2 _3} \right).
\label{X5}
\eea
To single out the leading singularity at the threshold in the interacting case, 
we use the
transformation formula\cite{grad},
\bea
&&F (a,b;c;z)\nn\\
&& = \frac{\Gamma(c) \Gamma(c-a-b)}{\Gamma(c-a)\Gamma(c-b)}
F(a,b;a+b-c+1;1-z)\nn\\
&&+  ~(1-z)^{c-a-b} \frac{\Gamma(c)\Gamma(a+b-c)}{\Gamma(a)\Gamma(b)}\nn\\
&& \times F(c-a,c-b;c-a-b+1;1-z)
\label{tr-formula}
\eea
and formally treat $1-z = 1 + {\bf q}^2/m^2 _3$ as a small parameter.
Then in the leading order
\bea
X^{xx}({\bf q}) &=& - \frac{1}{\pi} \frac{\Gamma(-2\theta)}{2^{1+2\theta}} (m_3\alpha)^{2\theta}
\left(\frac{\omega^2 _n - q^2}{m^2 _3} \right)\nn\\
&& \times \left( 1 + \frac{{\bf q}^2}{m^2 _3} \right)^{2\theta}
\left[ 1 + O \left( 1 + \frac{{\bf q}^2}{m^2 _3} \right)\right] 
\label{X6}
\eea
Performing analytical continuation ($\ri \omega_n \to \omega + \ri \delta$)
\bea
&&\left(\omega^2 _n + q^2 + m^2 _3  \right)^{2\theta}\nn\\
&&\to |s^2 - m^2 _3|^{2\theta} \left[ \theta(m^2 _3 - s^2) + \theta(s^2 - m^2 _3)
\cos 2\pi\theta \right]\nn\\
&& - \ri |s^2 - m^2 _3|^{2\theta}\theta(s^2 - m^2 _3)\sin 2\pi\theta~ 
{\rm sign}\omega
\label{analyt-cont}
\eea
and using the relation
\be
\Gamma(-2\theta) \sin 2\pi \theta = - \frac{\pi}{\Gamma(1+2\theta)},
\label{relation-gamma}
\ee
we arrive at the expression (\ref{LL-threshold}) of section \ref{sssec:TSF}.



\begin{thebibliography}{99}
\bibitem{comment} The expression (\ref{string-op-cont}) for the string operator
differs from that given in Refs.\cite{snt,nt,gnt} in that it is generalized to
the case of a finite coupling between the Ising copies.
\bibitem{prefactors} We use a different normalization of the fields as compared to
the one conventionally assumed in conformal field theory\cite{difran}. The OPEs given in this
Appendix transform to those of Ref.\cite{difran} under replacements:
$\xi \to \xi/\sqrt{2\pi}$, $\vare \to \vare/2\pi$,$\s \to 2^{-1/4} \s$, $\mu \to 2^{-1/4}\mu$.
\end{thebibliography}
\end{document}